\documentclass[preprint,aps,showpacs]{revtex4}

\input epsf

\def\Journal#1#2#3#4{{#1} {\bf #2}, #3 (#4)}

\def\NPB{{Nucl. Phys.} B}
\def\NPA{{Nucl. Phys.} A}
\def\PLB{{Phys. Lett.}  B}
\def\PRL{Phys. Rev. Lett.}
\def\PRD{{Phys. Rev.} D}
\def\PRC{{Phys. Rev.} C}

\def\JCP{J. Comp. Phys.}

\def\APPB{{Acta. Phys. Pol.} B}

\begin{document}

\title{Transport rates and momentum isotropization of gluon matter in 
ultrarelativistic heavy-ion collisions}

\author{Zhe Xu$^{1}$ \footnote{E-mail: xu@th.physik.uni-frankfurt.de}
and Carsten Greiner $^{1}$}
\affiliation{$^1$Institut f\"ur Theoretische Physik, Johann Wolfgang 
Goethe-Universit\"at Frankfurt, Max-von-Laue-Str.1, 
D-60438 Frankfurt am Main, Germany}

\date{March 2007}

\begin{abstract}
To describe momentum isotropization of gluon matter produced in
ultrarelativistic heavy-ion collisions, the transport rate of gluon drift
and the transport collision rates of elastic ($gg \leftrightarrow gg$) as
well as inelastic ($gg \leftrightarrow ggg$) perturbative quantum
chromodynamics- (pQCD) scattering processes are
introduced and calculated within the kinetic  parton cascade Boltzmann 
approach of multiparton scatterings (BAMPS), which simulates the
space-time evolution of partons. We define isotropization as the development
of an anisotropic system as it reaches isotropy. The inverse of the
introduced total transport rate gives the correct time scale of the momentum
isotropization. The contributions of the various scattering processes to
the momentum isotropization can be separated into the transport collision
rates. In contrast to the transport cross section, the transport collision
rate has an indirect but correctly implemented relationship with the
collision-angle distribution. Based on the calculated transport collision
rates from BAMPS for central Au+Au collisions at Relativistic Heavy Ion
Collider energies, we show that pQCD $gg \leftrightarrow ggg$ bremsstrahlung
processes isotropize the momentum five times more efficiently than elastic
scatterings. The large efficiency of the bremsstrahlung stems mainly from
its large momentum deflection. Due to kinematics, $2\to N$ $(N>2)$
production processes allow more particles to become isotropic in momentum
space and thus kinetically equilibrate more quickly than their back
reactions or elastic scatterings. We also show that the relaxation time in
the relaxation time approximation, which is often used, is strongly momentum
dependent and thus cannot serve as a global quantity that describes kinetic
equilibration.
\end{abstract}


\maketitle

\section{Introduction}
\label{intro}
It is speculated that the quark gluon plasma (QGP) created in 
Au+Au collisions at the Relativistic Heavy Ion Collider (RHIC) is 
a strongly coupled liquid \cite{GML05}. Because of strong coupling or
rather strong interactions, the QGP fluid has a very small viscosity.
However, questions regarding the source of the strong coupling
and its needed strength to generate a quasi-ideal fluid remain
unanswered. The necessary condition for the onset of perfect hydrodynamical
expansion is the achievement of local kinetic equilibrium. Although
the quarks and gluons produced at RHIC are far from thermal equilibrium,
kinetic equilibration should occur on a short time scale so that the
elliptic flow, $v_2$, increases substantially \cite{H01,PHEN04,STAR04-1}.
In this article we assume that the strong coupling and thermalization are
a consequence of frequent collisions among gluons on a semi-classical level.
We recently developed a new on-shell parton cascade code, BAMPS
(Boltzmann approach of multiparton scatterings) \cite{XG05}, which is
a microscopical relativistic transport model that solves the Boltzmann
equation for partons that are produced in ultrarelativistic heavy-ion
collisions. The included interactions can be elastic $gg \leftrightarrow gg$
processes or inelastic $gg \leftrightarrow ggg$ pQCD bremsstrahlung processes. 
Although the total perturbative quantum
chromodynamics- (pQCD) scattering cross section is only a few mb, it is
enough to drive the system toward full thermal equilibrium \cite{XG05}
and also to generate sufficiently large elliptic flow $v_2$ \cite{Xu2}.
Our goal is to understand the theoretical mechanism for the
fast equilibration of gluons, which are observed numerically.

In kinetic theory there are two competing processes that affect 
kinetic equilibration. The first is when particles stream freely 
between two subsequent collisions. In an expanding system free streaming
drives the system out of equilibrium. This is the case in a
one-dimensional Bjorken expansion, which most likely occurs early on
in ultrarelativistic heavy-ion collisions. The second one involves
collisions that make the particle momentum kinetically
isotropic and thermal. Here one has to take into account the
distribution of collision angle because large-angle collisions should 
contribute more to momentum isotropization. We define isotropization
as the development of an anisotropic system as it reaches isotropy.
A {\em transport cross section} \cite{DG85,T94} was introduced, either in the form
\begin{equation}
\label{tcs1}
\sigma^{{\rm tr}}=\int d\theta \frac{d\sigma}{d\theta} \sin^2\theta
\end{equation}
or
\begin{equation}
\label{tcs2}
\sigma^{{\rm tr}}=\int d\theta \frac{d\sigma}{d\theta} (1-\cos\theta)\,,
\end{equation}
where $\theta$ denotes the collision angle as a pertinent quantity that
measures the contributions of various collision processes to kinetic
equilibration. Although kinetic equilibration is observed locally in
the comoving frame of the expanding system, the transport cross
section is usually calculated in the center-of-mass (c.m.) frame of
individual colliding particles. The changes in momenta after the collision
appears different in each respective frame. Therefore,
the transport cross section may not be fully appropriate for
characterizing kinetic equilibration.

A widely used, yet simpler, method to characterize kinetic equilibration
is to calculate or estimate the relaxation
time $\tau_{{\rm rel}}$ \cite{baym,HK85,G91,HW96,W96,DG00,SS01}.
In the relaxation time approximation the collision term is expressed by
$(f_{{\rm eq}}-f)/\tau_{{\rm rel}}$, where $\tau_{{\rm rel}}$ is assumed
to be momentum independent and is then a global quantity that
characterizes the kinetic equilibration time scale. However, the validity
of the approximation must be verified.

In this article we derive a mathematical method of
quantifying the contributions of various processes to the momentum
isotropization. For this we define the {\em transport rate}, which 
is the momentum average of the particle density $f(x, p)$. The particle
density is found within the parton cascade as a solution of the Boltzmann
equation. Moreover, it will be shown that the inverse of
the total transport rate gives the global time scale of momentum
isotropization. In Sec. \ref{sec:1} we mention the operation of
the employed parton cascade BAMPS and improvements made in it. 
The initial condition of gluons, as an input for the parton cascade, is 
discussed in Sec. \ref{sec:2}. We show results on thermal equilibration
and momentum isotropization of gluons in Sec. \ref{sec:3} for a central
Au+Au collision at RHIC ($\sqrt s=200$ GeV). The inclusion of quarks
into the parton cascade is straightforward and the results will be presented 
in another article. In Sec. \ref{sec:4} we define
the transport rates, which determine contributions of various processes to
the momentum isotropization and derive their relations to
the transport cross sections. We present in Sec. \ref{sec:5}
the numerical results on the transport rates. The transport rate of gluon
drift is computed and compared with the one when obtained assuming Bjorken
boost invariance. To show the importance of the bremsstrahlung
processes in thermal equilibration, we carry out simulations with and
without these processes for comparison. The quantitative difference
in the momentum isotropization for both simulations is manifested by
the ratio of the total transport rates. The ratio, which turns out to
be approximately 6, is used to perform a third type of simulation in
which only elastic scatterings with artificially enlarged cross sections
are included. Although such large cross sections are not physical,
they verify our main finding: the total transport collision
rate is the key quantity determining momentum isotropization.
Despite the process type as long as the total transport collision rate
is the same, the momentum isotropization is also the same.
At the end of Sec. \ref{sec:5} we demonstrate that the relaxation time
approximation is not suitable for the quantification of the time scale
for kinetic equilibration. A summary of our findings is given in
Sec. \ref{summary}. Detailed expressions for calculating the transport
rates are derived in Appendix \ref{app1}.

\section{BAMPS and Setup}
\label{sec:1}
The structure of the parton cascade BAMPS is based on the stochastic
interpretation of the transition rate \cite{DB91,L93,C02,XG05}.
This interpretation ensures that detailed balance is not violated, which
is nontrivial when the geometrical concept of cross section is used
\cite{M99}, especially for multiple scatterings like
$ggg \leftrightarrow gg$. BAMPS subdivides space into small cell units.
In each of which we separately evaluate the transition probabilities of
all possible gluon pairs and triplets to see if a particular scattering
(or transition) occurs. The smaller the cells the more local transitions
can be realized. However, the smaller cells contain fewer
particles and thus have larger statistical fluctuations in their 
calculated transition rates. To achieve a high-enough number of
pairs and triplets of gluons in a cell, we adopt a test particle technique,
which amplifies the (pseudo)gluon density by a factor of $N_{{\rm test}}$. 
Accordingly, the cross sections have to be reduced by the same factor to
obtain the same physical mean free path \cite{XG05}. In this article
the transverse length of a cell is a constant of $\Delta x=\Delta y=0.25$ fm
and the longitudinal length $\Delta z$ is half of that in \cite{XG05},
so for a cell of the center of the collision $\Delta z\approx 0.1t$, where
$t$ is the running time of the evolution of gluon matter. $N_{{\rm test}}$
is set to $280$, which ensures that there are on average $15$ test particles
per cell.

The differential cross section for the elastic pQCD scatterings of
gluons is given by
\begin{equation}
\frac{d\sigma^{gg\to gg}}{dq_{\perp}^2} =
\frac{9\pi\alpha_s^2}{(q_{\perp}^2+m_D^2)^2}\,.
\end{equation}
Three-body gluonic interactions are described by the effective matrix
element
\cite{GB82,biro,W96}
\begin{equation}
\label{m23}
| {\cal M}_{gg \to ggg} |^2 = \frac{9 g^4}{2}
\frac{s^2}{({\bf q}_{\perp}^2+m_D^2)^2}\,
 \frac{12 g^2 {\bf q}_{\perp}^2}
{{\bf k}_{\perp}^2 [({\bf k}_{\perp}-{\bf q}_{\perp})^2+m_D^2]}
\, \Theta(k_{\perp}\Lambda_g-\cosh y)
\end{equation}
where $g^2=4\pi\alpha_s$. $\alpha_s$ is set to $0.3$ in contrast
to the running coupling used in Ref. \cite{XG05}. ${\bf q}_{\perp}$ and
${\bf k}_{\perp}$ denote the perpendicular component of the momentum
transfer and of the radiated gluon momentum in the center-of-mass
frame of the collision, respectively. $y$ is the momentum rapidity of
the radiated gluon in the center-of-mass frame, and $\Lambda_g$ is the
mean free path of a gluon.

We regularize the infrared divergences by introducing
the Debye screening mass $m_D$
\begin{equation}
\label{md2}
m_D^2 = 16\pi \alpha_s N_c \int \frac{d^3p}{(2\pi)^3} \frac{1}{p} \,f_g
\end{equation}
($N_c=3$), which is calculated locally using the current gluon density
obtained from BAMPS. In general, the Debye screening mass should
depend on the direction of the gluon propagator \cite{BMW92}. If
the gluon distribution $f_g$ significantly deviates from its isotropic
shape, the Debye screening mass may even become negative, which leads to 
instabilities in certain modes of the soft gauge field 
\cite{M93,ALM03,A05,R05,D05,SSGT06}. These instabilities and their proper
inclusion are beyond the scope of the present article. We have simplified
the problem by removing the directional dependence of the Debye
screening mass.

The suppression of the radiation of soft gluons due to
the Landau-Pomeranchuk-Migdal (LPM) effect \cite{biro,W96,XG05}
is included using the step function in Eq. (\ref{m23}). There the time of
the emission, $\sim \frac{1}{k_{\perp}} \cosh y$, should be smaller than
the time interval between two scatterings or equivalently the gluon mean
free path $\Lambda_g$. This leads to a lower cutoff for $k_{\perp}$ and
a decrease in the total cross section or the transition probability.

Compared to the default setup in Ref. \cite{XG05}, further improvements
have been made. To calculate the Debye screening mass $m_D$ in a local
region more accurately, we make use of the polar symmetry in central
collisions and divide the transverse plane in each $\Delta z$-bin into
rings: the first ring has a radial size of $0 < x_T < 1.5$ fm ($x_T$ being
the transverse radius), and the following rings have transverse radial
widths of $1$ fm. The rings are regarded as local regions in which the Debye
screening mass is evaluated.

The local collision rates of all interaction channels, the sum of which
is the inverse of the mean free path that models the LPM effect, were
evaluated in Ref. \cite{XG05} in individual cells. This leads
to large fluctuations in the mean free path in cells with few
(test) particles. To reduce these fluctuations we
take the averaged value of the collision rates over all the cells within
individual rings. In addition, transverse velocities of rings are taken
into account to calculate the collision rates in the comoving frames.

Moreover, we assume that if the energy density, which
is calculated locally in the comoving frame, sinks below
$1 \mbox{ GeV/fm}^3$, particles in that region no longer interact,
so they propagate freely. At this stage a hadronization procedure
should be applied, which is planned as a future project.

We concentrate on the central region of the full reaction, which is
defined as a cylinder with $0 < x_T < 1.5$ fm and $-0.2 < \eta < 0.2$
where $\eta$ denotes the space-time rapidity
$\eta=\frac{1}{2}\ln \frac{t+z}{t-z}$. The longitudinal extension of
the cylinder is, thus, $\Delta z = 2\,t\,\tanh(0.2)\approx 0.4\,t$.
The parameters for bounding the cylinder are found by balancing between
having a small, local region and avoiding high statistical fluctuations.
Results in this region are obtained by averaging over the various
ensembles.

\section{Initial conditions}
\label{sec:2}
Initial gluons are taken as an ensemble of minijets with transverse
momentum greater than $1.4$ GeV, which are produced via semihard
nucleon-nucleon collisions \cite{EKL89}. Details of the
distribution of the initial gluons in space and time can be found in
Ref. \cite{XG05}. Using Glauber-geometry and assuming independent binary
nucleon-nucleon collisions, the gluon number is initially about $700$
per momentum rapidity. These gluons take about $60\%$ of the total
given energy entered in a central Au+Au collision. The lower momentum
cutoff is taken as a parameter to fit the experimentally measured
final transverse energy at midrapidity (see Fig. \ref{et}).

For simplicity's sake one may assume that the two gold nuclei are
extremely Lorentz contracted with zero width. Assuming that on-shell
gluons are immediately formed (i.e., without any formation time) at the same
time when the corresponding nucleon-nucleon collision occurs, all initial
gluons are positioned at $z=0$ fm at $t=0$ fm/c. Subsequent free
streaming would immediately order the gluons with the momentum rapidity $y$
to a spatial slice with the space-time rapidity $\eta$ being equal to $y$.
In the comoving frame of each spatial slice gluon momentum has only 
a transverse component and it has a highly anisotropic distribution. 

\begin{figure}[ht]
\centerline{\epsfysize=7cm \epsfbox{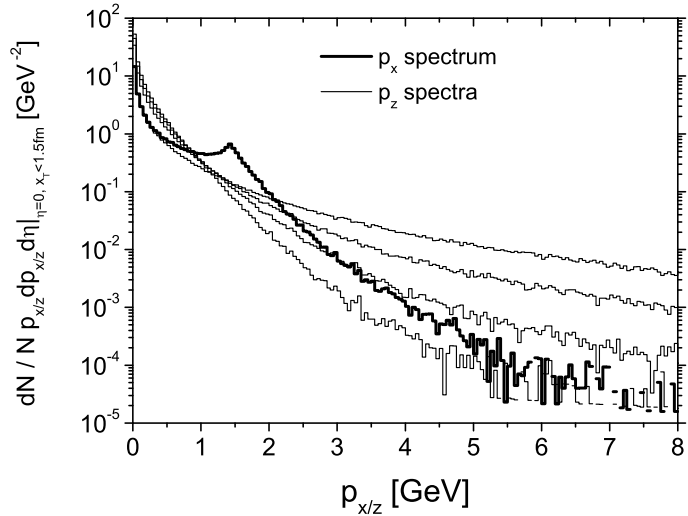}}
\caption{Transverse and longitudinal gluonic momenta spectra
during initial free streaming. The thick histogram is the
distribution of $|p_x|$ at a time of $0.1$ fm/c, whereas the
thin histograms are the distributions of $|p_z|$ at times of
$0.1$, $0.2$, $0.22$, and $0.24$ fm/c, respectively, from top
to bottom. The results are obtained from the central region.
}
\label{initdistpxz}
\end{figure}

At RHIC energy each of the colliding gold nuclei has a small but
nonvanishing longitudinal extension of about $0.2$ fm. Therefore, gluons
are primarily produced at $z=0$ fm (or $\eta=0$) at $t\simeq 0.1$ fm/c
when the two nuclei overlap fully. Note that $t=0$ fm/c is when two nuclei
are just touching. In contrast to the simplified case mentioned above 
in reality there is a significant smearing in the gluonic $\eta-y$
correlation for times $t\le 0.2$ fm/c. For instance, gluons with $y \ne 0$
will also appear in the central slice with $\eta=0$ for a while.
The rate of smearing disappearance is shown in Fig. \ref{initdistpxz},
where the spectra of transverse and longitudinal gluon momenta are depicted
during initial free streaming in absence of secondary collisions.

The spectra are obtained in the central region ($0 < x_T < 1.5$ fm and
$-0.2 < \eta < 0.2$). We see that the $|p_z|$ spectrum changes quite 
drastically after $0.2$ fm/c, the same point when two gold nuclei
cease to overlap and the production of minijets is completed.
The free streaming of high $|p_z|$ gluons away from the central region
leads to strong, continuous suppression in the $|p_z|$ spectrum.
Corresponding to this suppression, the changes in the $|p_x|$ spectrum
at low transverse momentum are, however, tiny compared with their absolute
values at low $|p_x|$. At high $|p_x|$ the change in time is negligable
because there is no initial transverse expansion for large nuclei.
Therefore, in Fig. \ref{initdistpxz} the $|p_x|$ spectrum
is depicted only at $t=0.1$ fm/c. We note that at large $|p_z|$ 
the suppression stops completely when all the particles with high $p_z$
(or with high $y > \eta_b=0.2$) have left the small, but finite, rapidity
window $[-\eta_b:\eta_b]$. Then, only particles with lower rapidity $y$
remain in the central region. The time a gluon needs to leave the central
region is, for instance,
$\Delta t=t_0\,\tanh \eta_b/(\tanh y-\tanh \eta_b)$ when the gluon is
produced at $z_0=0$ fm at $t_0$. We see that the larger the momentum
rapidity $y$, the smaller $\Delta t$. For a gluon with $y=1$ and
$t_0=0.1$ fm/c it takes $\Delta t=0.035$ fm/c to leave the central region.

Comparing the particle spectrum of
$|p_z|$ with that of $|p_x|$, the momentum distribution is, strictly
speaking, at no time isotropic during the initial free streaming.
The characteristic hump in the $|p_x|$ spectrum at $1.4$ GeV arises
from the requirement that the transverse momentum of the original 
minijets should be greater than $1.4$ GeV. 
Choosing other initial conditions like in HIJING \cite{WG91,EW94} or 
the color glass condensate \cite{MV94} would change the shape of the initial 
momentum distribution. However, even though the momentum distribution 
might be isotropic during the continuing suppression of high $|p_z|$ gluons,
the further suppression leads to a deviation in the momentum distribution
away from isotropy within a very short time of $\sim 0.1$ fm/c.
Therefore, we can conclude that free streaming leads to
$\eta\approx y$ regardless of the initial $\eta-y$ correlation.
The gluon momentum distribution after short-time free streaming
is, in general, neither thermal nor isotropic.

In this article we introduce an
additional {\it formation time} \cite{XG05} for every minijet,
$\Delta t_f=\cosh y \Delta \tau_f\approx \cosh y\cdot 1/p_T$, which
models the prior off-shell propagation of the gluons to be freed
in individual nucleon-nucleon collisions, where $\cosh y$ denotes
the Lorentz factor. Within $\Delta t_f$ we assume that the virtual gluon
does not interact and, therefore, moves freely at the speed of light.
Gluons with large $|p_z|$ have in turn a large Lorentz factor and, thus,
a large formation time. Although most of these gluons are produced in
the central region, they are far from the central region when they materialize
as on-shell partons because of the assumed off-shell propagation.
Because we count particles only if they are on-shell, i.e., interactive,
the initial gluon momentum distribution at $0.2$ fm/c differs
from that shown in Fig. \ref{initdistpxz}. However, it is practically
identical to that at $0.24$ fm/c, which is not isotropic.
When the initial conditions are chosen accordingly and a simulation
including the pQCD bremsstrahlung processes is performed, we obtain
$dE_T/dy \approx 640$ GeV at midrapidity with a final time of $5$ fm/c,
at which the energy density of gluons decreases to the critical value of 
$1$ GeV/$\mbox{fm}^3$. Our $dE_T/dy$ at $y=0$ from the simulation is
comparable with that found in experimental measurements
at RHIC \cite{STAR04-2} (see Fig. \ref{et}).

\section{Momentum isotropization and kinetic equilibration}
\label{sec:3}
Kinetic equilibration is a process in which the particle momentum becomes
isotropic and thermal, which has an exponential distribution. Momentum
isotropization is part of kinetic equilibration
and is reached before full kinetic equilibrium \cite{A05,R05,D05}.
(Strictly speaking, full kinetic equilibrium can be achieved only for
a static, nonexpanding system.)
In this article we concentrate on the contribution of collision processes
to momentum isotropization and kinetic equilibration of gluon matter
in ultrarelativistic heavy-ion collisions.

As demonstrated in Sec. \ref{sec:2}, the initial free streaming (or 
the off-shell propagation) of gluons with high momentum rapidity $y$
makes the momentum distribution anisotropic, even if it appears
momentarily isotropic. Initially in the central region most gluons move in
the transverse direction. Secondary collision processes gradually force
them into the longitudinal direction, which gives
a positive contribution to momentum isotropization. However,
whenever a gluon switches to the longitudinal direction, its momentum
rapidity grows and the gluon tends to drift out of the central region.
This gives a negative contribution to momentum isotropization in the local
region. Although gluons with the same (regardless of $\pm$ sign) momentum 
rapidity drift from their neighboring slices into the central slice,
this cannot completely compensate for the loss in the central region. 
The reason is that thermalization occurs earlier in the central slice 
than in the outwards regions corresponding to Bjorken's picture of 
boost-invariance in the space-time evolution of the parton system \cite{B83}.
In the transverse direction, however, there is no transverse flow 
at the beginning of the expansion. Therefore, no net drift of gluons 
occurs in the transverse direction. The difference in the gluon drift
in the longitudinal and transverse directions leads to a situation in which
the net effect of the drift has a negative contribution to momentum
isotropization and the stronger the momentum isotropization, the larger
the negative contribution of the particle drift. At later times, when
three-dimensional expansion takes place, there is also a net particle
drift in the transverse direction and the negative contribution of
the particle drift to momentum isotropization decreases.

\begin{figure}[ht]
\centerline{\epsfysize=7cm \epsfbox{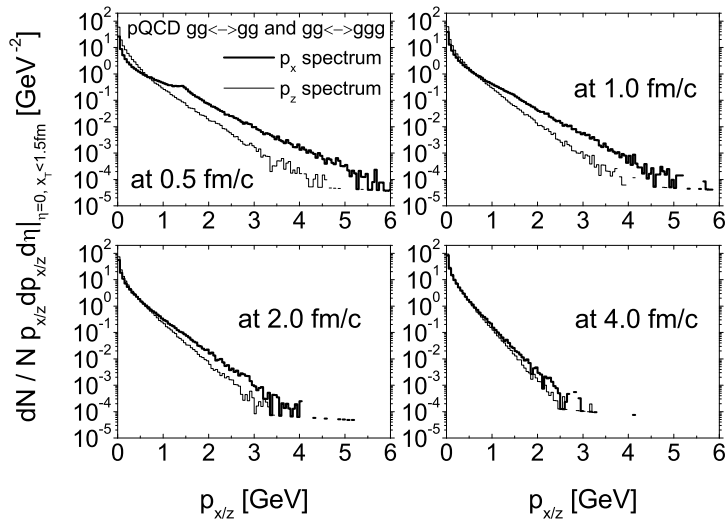}}
\caption{Transverse (thick histograms) and longitudinal
(thin histograms) gluon momentum spectra in the central region at
various times throughout the evolution of gluon matter. Results are
obtained from the simulation when both elastic and inelastic 
pQCD-scattering processes are included.
}
\label{distpxz}
\end{figure}

In this section we first demonstrate momentum isotropization and kinetic
equilibration of gluons in central Au+Au collisions at RHIC energy.
The various contributions of collisions and drift to momentum
isotropization will be analysed in detail in the next section.
Figure \ref{distpxz} depicts the transverse and longitudinal gluon
momenta distributions in the central region at four
different times throughout the evolution of gluon matter.
The results are obtained when both elastic and inelastic pQCD-based
scattering processes are included. We first see that the momentum
distribution continuously isotropizes and thermalizes over time.
Due to the expansion full thermal equilibrium cannot be achieved if
the collision rate is finite. A certain mismatch in $p_z$ and $p_x$ 
must exist due to the counteraction between the expansion and the collisions.

Second, Fig. \ref{distpxz} shows that there is an exponential
distribution before the system becomes isotropic. It is almost impossible
to distinguish the gluon momentum isotropy time scale from the
thermal time scale. It seems that when collisions drive the particle
momentum close to isotropy, the momentum distribution is already 
practically thermalized. In general, momentum isotropization happens on
a shorter time scale than kinetic equilibration. The difference in
the time scales of both dynamical processes depends on the initial
condition for gluons.

Kinetic equilibration for the softer gluons is completed earlier than
that for the harder gluons. This is obvious for elastic
$gg\leftrightarrow gg$ scattering processes because the momentum transfer
in collisions is typically the Debye screening mass. Therefore,
the hard gluons cannot be deflected as strongly as the soft gluons.
However, in the inelastic pQCD $gg\leftrightarrow ggg$
collisions, which we will prove are the dominant processes in kinetic
equilibration, the difference in the momentum degradation for soft
and hard gluons is miniscule due to the production or absorbtion of an
additional gluon. Averaging the various kinetic equilibrium times
for soft and hard gluons the momentum distribution
becomes {\em isotropic and thermal at $1-2$ fm/c}. Furthermore, we clearly
see that the distributions become steeper with time, which 
indicates the ongoing cooling of the system related to
quasihydrodynamical behavior due to the subsequent work done by the
expanding system.

\begin{figure}[ht]
\centerline{\epsfysize=7cm \epsfbox{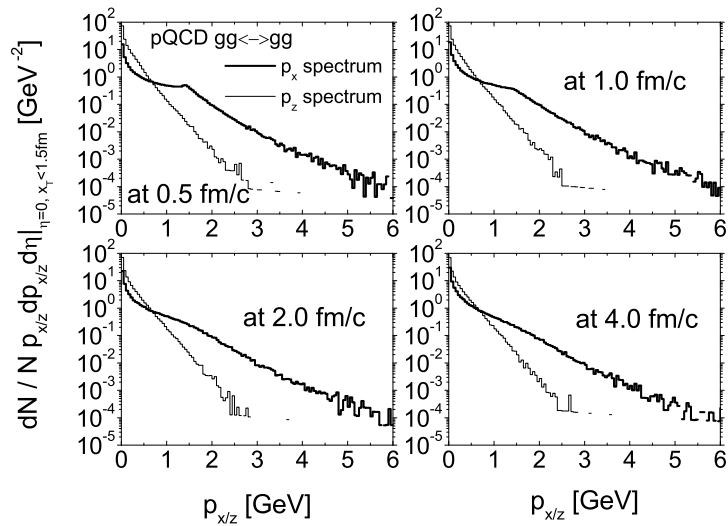}}
\caption{Transverse (thick histograms) and longitudinal
(thin histograms) gluon momentum spectra in the central region at
various times throughout the evolution of gluon matter. Results are
obtained using elastic only pQCD-scattering processes.
}
\label{p22distpxz}
\end{figure}

To understand the role of the inelastic pQCD
$gg\leftrightarrow ggg$ processes in kinetic equilibration, we
also carry out calculations in which gluons interact only via
elastic scatterings. The initial conditions are the same as those
when inelastic collisions are included. The results
are shown in Fig. \ref{p22distpxz}, which has the same structure
as Fig. \ref{distpxz}. 
The difference in the results depicted in both figures can be
immediately seen. The spectra in Fig. \ref{p22distpxz} only show a small 
change throughout the entire evolution of the system and are still
highly anisotropic and are not thermalized as late as $4$ fm/c.
The evolution resembles that of free streaming.

\begin{figure}[ht]
\centerline{\epsfysize=7cm \epsfbox{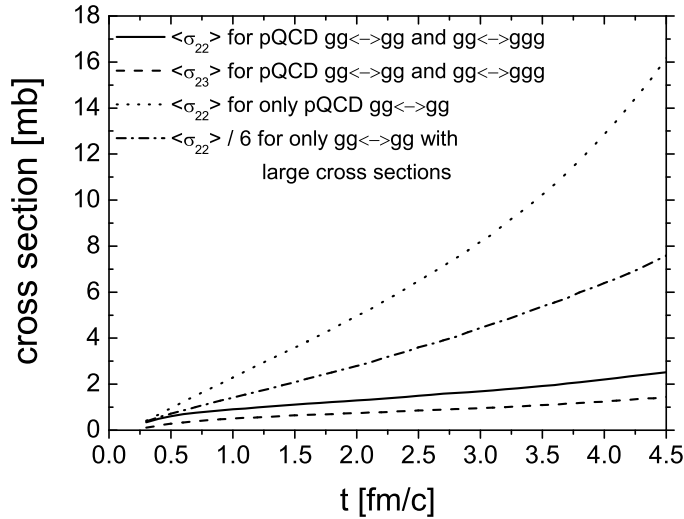}}
\caption{Time evolution of pQCD cross sections. The solid
and dashed curve show the pQCD cross section for
$gg\to gg$ and $gg\to ggg$ collisions, respectively, when both
elastic and inelastic scattering processes are considered. The dotted
curve gives the cross section of $gg\to gg$ collisions in the simulation,
including elastic-only pQCD scattering processes, whereas the 
dash-dotted curve indicates the cross section ({\em divided by a factor of 6})
of $gg\to gg$ collisions when elastic-only scattering processes with
artificially large cross sections are included.
}
\label{cross-section}
\end{figure}

The kinetic equilibration time strongly depends on whether the pQCD
bremsstrahlung processes and their back reactions are
taken into account. The pQCD bremsstrahlung processes and their
back reactions isotropize the momentum more efficiently than elastic
collisions and, thus, play an essential role in early
thermalization of gluons in heavy-ion collisions at RHIC.
As seen in Fig. \ref{cross-section}, the pQCD cross 
section of $gg\to ggg$ processes, including LPM suppression 
(dashed curve), is {\em smaller} than that of elastic scatterings 
(solid curve) and much smaller than the cross section obtained in 
the simulation with elastic-only scattering processes (dotted curve).
Although particle production in inelastic processes can enhance
the number of collision centers and, thus, effectively shorten the mean
free path of particles, chemical equilibration will balance the
production by the annihilation of particles to avoid oversaturation.
The fact that the cross section of the pQCD bremsstrahlung process
is small, but its kinetic equilibration efficiency is large,
demonstrates that cross sections or collision rates are not the correct
quantities to describe the contributions of various processes to
kinetic equilibration. The  collision-angle distribution must
be at least taken into account. Defining the correct quantity is one of
the main purposes for this article.

In Fig. \ref{cross-section} the large difference in the total cross
sections of elastic scatterings for various simulations is shown. Because
\begin{equation}
\langle \sigma_{gg\to gg} \rangle \, \sim \, \frac{1}{m_D^2\,
\langle 1+4m_D^2/s\rangle }\,,
\end{equation}
the difference in the cross sections arises from the difference in the
development of the Debye screening mass $m_D$ in the various simulations.
$m_D$ is calculated dynamically according to (\ref{md2}) and, thus,
is roughly proportional to $\sqrt{n/\langle p\rangle}=n/\sqrt{\epsilon}$,
where $n$ and $\epsilon$ are the number and energy density of
gluons, respectively. 

We consider two extreme cases of expansion with initial conditions that
possess the longitudinal boost invariance. One case is free streaming,
for which $n$ as well as $\epsilon$ decrease as $\tau^{-1}$, where
$\tau=\sqrt{t^2-z^2}$ is the proper time. Thus, $m_D$ decreases as
$\tau^{-1/2}$. In the other case of a one-dimensional ideal hydrodynamical
expansion, $n$ decreases as $\tau^{-1}$, whereas $\epsilon$ decreases
as $\tau^{-4/3}$. Therefore, $m_D$ decreases as $\tau^{-1/3}$. In a viscous
hydrodynamical expansion the decrease of $m_D$ over time falls between
the two cases. The time dependence of the Debye screening mass in a real
expansion starting out of thermal equilibrium and undergoing thermalization
is more complicated. Whereas kinetic equilibration drives the density 
distribution of gluons to its thermalized shape, which affects the calculation
of $m_D$ (\ref{md2}), chemical equilibration, which is not 
taken into account above, will enhance or reduce the gluon
number, which in turn enhances or reduces $m_D$.

\begin{figure}[ht]
\centerline{\epsfysize=7cm \epsfbox{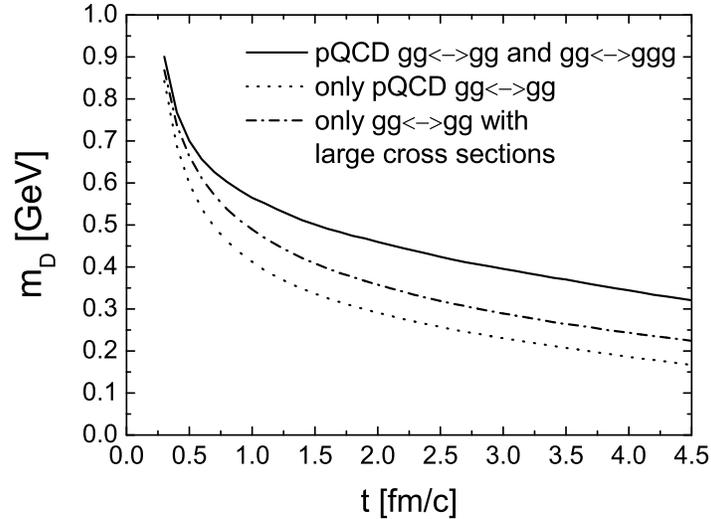}}
\caption{Time evolution of the Debye screening mass. Results are obtained
from the simulation with both pQCD elastic and inelastic collisions (solid
curve), with elastic-only pQCD collisions (dotted curve), and with
elastic-only collisions using large cross sections (dash-dotted curve).
}
\label{md}
\end{figure}

\begin{figure}[ht]
\centerline{\epsfysize=7cm \epsfbox{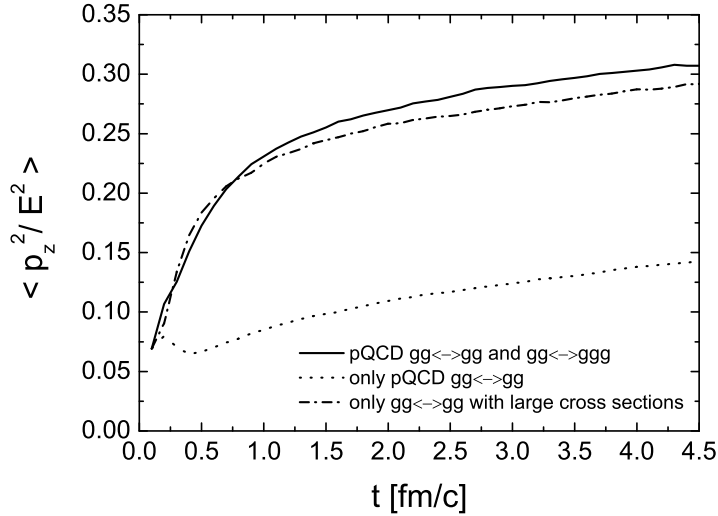}}
\caption{
Momentum isotropization. Results are obtained from the simulation with
both elastic and inelastic pQCD-scattering processes (solid curve),
with elastic-only pQCD-scattering processes (dotted curve) and with
elastic-only scattering processes using artificially large cross sections
(dash-dotted curve).
}
\label{pze2}
\end{figure}

Figure \ref{md} shows the time evolution of the Debye screening mass
in various simulations.
The results are obtained in the central region where $t\approx \tau$.
All the curves in Fig. \ref{md} decrease with time. Similar calculations
for the Debye screening mass have been done in Refs. \cite{SS00, BMS03} 
employing the parton cascade VNI/BMS. From Fig. \ref{md} we see that
the evolution depends on the simulation type: in the simulation including both
elastic and inelastic pQCD scatterings (solid curve) $m_D$ decreases
slower than $t^{-1/3}$ due to gluon production in the
course of chemical equilibration; in the simulation with elastic-only
pQCD collisions the decrease of $m_D$ (dotted curve) is 
slightly stronger than $t^{-1/2}$, which indicates again that the 
expansion of gluons in this simulation resembles that of free 
streaming; the third simulation includes elastic-only scatterings 
with artificially large cross sections and shows the same kinetic
equilibration as that in the simulation including both elastic and
inelastic pQCD collisions (see Fig. \ref{pze2}). The Debye screening
mass in this simulation (dash-dotted curve) decreases between $t^{-1/2}$
and $t^{-1/3}$.

Returning to the kinetic equilibration analysis, the gluon
kinetic equilibration time can in principle be determined quantitatively
by studing the entropy production. Because the entropy can be hardly
extracted from any microscopic cascade, we concentrate on momentum
isotropization of gluons. Choosing minijets production as the initial
condition, momentum isotropization and kinetic equilibration time scales
are almost identical (as seen in Fig. \ref{distpxz}).

To quantify momentum isotropization we have to choose an appropriate
momentum-distribution moment $Q$. For instance, $Q:=\langle p_z^2/E^2\rangle$
is used to describe momentum isotropization. Later we briefly discuss
the consequences of $Q=\langle |p_z|/E\rangle$,
to see how sensitive the results are to different descriptions
of momentum isotropization. In Fig. \ref{pze2} momentum isotropization
with $Q=\langle p_z^2/E^2\rangle$ is depicted.
The average is taken over all gluons in the central region. In
Fig. \ref{pze2} we see that $Q$ relaxes its equilibrium value of $1/3$
when the inelastic processes are included, whereas it still deviates from
its equilibrium value at the time $4.5$ fm/c
when only elastic pQCD scatterings are considered. These results agree
with the momentum spectra time evolution shown in Figs. \ref{distpxz} and
\ref{p22distpxz}. The dash-dotted curve in Fig. \ref{pze2} depicts
the momentum isotropization considering elastic-only collisions with 
artificially large cross sections and is almost the same as the
solid curve, which implies they have  basically the same kinetic
equilibration. The third simulation is detailed in Sec. \ref{sec:5}.

The momentum isotropization fit is found using the relaxation formula
\begin{equation}
\label{fit}
F(t)=\frac{1}{3}+ \left [ Q (t_0)
-\frac{1}{3} \right ] \exp\, \left [-\frac{t-t_0}{\theta_{{\rm rel}}(t_0)}
\right ]\,.
\end{equation}
$F(t)$ is equal to $Q(t)$ only at $t=t_0$.
For every fixed $t_0$ the relaxation time $\theta_{{\rm rel}}$ is constant
with respect to $t$. Using $\theta_{{\rm rel}}=0.9$ fm/c at $t_0=0.3$ fm/c
up to $1.0$ fm/c and $\theta_{{\rm rel}}=2.4$ fm/c at $t_0=1.2$ fm/c for
the rest, $F(t)$ is a perfect fit for the solid curve in Fig. \ref{pze2}.
An isotropy is achieved at about $1.0$ fm/c in the simulation that 
includes both elastic and inelastic pQCD-scattering processes. This
time scale is consistent with that extracted from the momentum distribution
(see Fig. \ref{distpxz}). Within our parton cascade description early
thermalization occurs at roughly $1$ fm/c for the initially
nonequilibrated gluon matter at RHIC.

The relaxation time $\theta_{{\rm rel}}$ is generally time dependent.
Because a local fit requires that the time derivatives of $F(t)$ and $Q(t)$
are equal at $t=t_0$, which leads to
\begin{equation}
\label{relax1}
\left . \dot Q(t) \right |_{t=t_0}=\left . \dot F(t) \right |_{t=t_0}
=-(Q(t_0)-Q_{{\rm eq}}) \, \frac{1}{\theta_{{\rm rel}}(t_0)}\,,
\end{equation}
where $Q_{{\rm eq}}=1/3$, $\theta_{{\rm rel}}$ can be calculated as
[changing $t_0$ to $t$ in Eq. (\ref{relax1})]
\begin{equation}
\label{relax2}
\frac{\dot Q(t)}{Q_{{\rm eq}}-Q(t)}=\frac{1}{\theta_{{\rm rel}}(t)}.
\end{equation}
Equation (\ref{relax2}) expresses the relaxation rate of momentum
isotropization $1/\theta_{{\rm rel}}$ as a function of time, which will
be separated analytically into terms corresponding to the particle drift
and the various scattering processes.  We will derive the so-called
{\it transport rates}, which {\it precisely} quantify the contributions of 
various processes to momentum isotropization.

\section{Transport rate}
\label{sec:4}
To introduce $Q$ at a certain space point $\vec \xi$ one has to consider
its comoving frame. For the coordinate $x$ and the momentum four-vector
$p$ in the $\vec \xi$'s comoving frame, $Q$ is defined by
\begin{equation}
\label{aniso}
Q(t):=\left. \left \langle \frac{p_z^2}{E^2} \right \rangle 
\right |_{\vec x=0}=
\frac{1}{n}\int \frac{d^3p}{(2\pi)^3}
\frac{p_z^2}{E^2} \, \left. f(\vec x, t, p) \right |_{\vec x=0}\,,
\end{equation}
where the local number density is
\begin{equation}
\label{ndens}
n(t)=\int \frac{d^3p}{(2\pi)^3} \, \left. f(\vec x, t, p) \right |_{\vec x=0}
\,.
\end{equation}
In practice $Q$ is evaluated within a volume element, which is small
compared to the volume of the expanding system
but is large enough so that it still contains a large number of particles.
For the calculations shown in Fig. \ref{pze2} we used the central region
bounded by $x_T < r_b=1.5$ fm and $|\eta|<\eta_b=0.2$.
Correspondingly (\ref{aniso}) and (\ref{ndens}) must also be adjusted and
detailed expressions are derived explicitly in Appendix \ref{app1}.
As a simplification we now consider the limit $r_b\to 0$ and $\eta_b\to 0$
where definitions (\ref{aniso}) and (\ref{ndens}) can be used.

Taking the time derivative of $Q(t)$ yields
\begin{equation}
\label{dotaniso}
\dot Q(t)=\frac{1}{n}\int \frac{d^3p}{(2\pi)^3}\frac{p_z^2}{E^2} \, 
\left. \frac{\partial f}{\partial t}\right |_{\vec x=0}
-Q(t)\frac{1}{n}\int \frac{d^3p}{(2\pi)^3}
\left. \frac{\partial f}{\partial t}\right |_{\vec x=0}\,.
\end{equation}
We replace $\partial f/\partial t$ in Eq. (\ref{dotaniso}) by
\begin{equation}
\label{boltzmann}
\frac{\partial f}{\partial t}=-\frac{\vec{p}}{E} \cdot \vec{\nabla} f +
C_{22}+C_{23}+C_{32}
\end{equation}
from the Boltzmann equation, where $-\frac{\vec{p}}{E} \cdot \vec{\nabla} f$
corresponds to particle drift and $C_{22}$, $C_{23}$, and $C_{32}$
denote the collision terms corresponding to $gg\to gg$,
$gg\to ggg$, and $ggg\to gg$, respectively. It is obvious that
the contribution of the various processes to $\dot Q(t)$ is additive.
We rewrite Eq. (\ref{dotaniso})
\begin{equation}
\label{dotaniso1}
\dot Q(t)=W_{{\rm drift}}(t)+W_{22}(t)+W_{23}(t)+W_{32}(t)\,,
\end{equation}
where $W_{{\rm drift}}$, $W_{22}$, $W_{23}$, and $W_{32}$ correspond to 
particle drift, $gg\to gg$, $gg\to ggg$, and $ggg\to gg$ collision
processes, respectively. According to Eq. (\ref{relax2}) we obtain
\begin{equation}
\label{separate}
\frac{1}{\theta_{{\rm rel}}(t)}=R^{{\rm tr}}_{{\rm drift}}(t)
+R^{{\rm tr}}_{22}(t)+R^{{\rm tr}}_{23}(t)+R^{{\rm tr}}_{32}(t)\,,
\end{equation}
where we define
\begin{equation}
\label{trate}
R^{{\rm tr}}_i(t):=\frac{W_i(t)}{Q_{{\rm eq}}-Q(t)}
\end{equation}
for $i={\rm drift}$, $22$, $23$, and $32$. One sees that the relaxation rate
of momentum isotropization $1/\theta_{{\rm rel}}$ is separated into
additive parts corresponding to the particle drift and the various
collision processes. $R^{{\rm tr}}_{\rm drift}$ is called the
{\em transport rate of particle drift}, whereas $R^{\rm tr}_{22}$, $R^{\rm tr}_{23}$,
and $R^{\rm tr}_{32}$ stand for the {\em transport collision rates} of
their respective interactions. Extending this to more than three-body
processes is straightforward because the collision term is additive.
We note that $R^{\rm tr}_i$ (shown below) depends on the definition
of $Q$. When one changes $Q$ from $Q=\langle p_z^2/E^2 \rangle$ to
$Q=\langle |p_z|/E\rangle$, the form of $R^{\rm tr}_i$ changes
accordingly.

\subsection{$R^{\rm tr}_{\rm drift}$}
Except for static systems the drift term in the Boltzmann
equation (\ref{boltzmann}) generally contributes to $\dot Q(t)$. 
$W_{\rm drift}$ is given by
\begin{equation}
\label{diff}
W_{\rm drift}(t)=\frac{1}{n} \int \frac{d^3p}{(2\pi)^3}\,
\frac{\vec p}{E}\cdot \vec \nabla f \, 
\left [ Q(t)-\frac{p_z^2}{E^2} \right ]\,.
\end{equation}
Assuming Bjorken's space-time picture of a central ultrarelativistic
heavy-ion collision \cite{B83}, we can use the relation
\begin{equation}
\label{baymrelat}
\frac{\vec p}{E} \cdot \vec \nabla f \approx 
\frac{p_z}{E} \frac{\partial f}{\partial z}
=-\frac{p_z}{t}\frac{\partial f}{\partial p_z}
\end{equation}
found in Ref. \cite{baym}. Inserting Eq. (\ref{baymrelat}) into (\ref{diff})
and performing partial integrals we obtain
\begin{equation}
\label{diff-2}
R^{\rm tr}_{\rm drift}(t) \approx \frac{-2}{\left [Q_{\rm eq}-Q(t)
\right ]\,t}\,
\left [ Q(t)-\left \langle \frac{p_z^4}{E^4} \right \rangle (t) \right ]\,.
\end{equation}
Equation (\ref{diff-2}) shows that $R^{\rm tr}_{\rm drift}$ is negative,
which agrees with our conclusion in the previous section. Using
the approximation $\langle p^4_z/E^4\rangle \approx Q^2$, we see
that the larger the Q, the larger the $-R^{\rm tr}_{\rm drift}$. 

\subsection{$R^{\rm tr}_{22}$}
Changing $p$ to $p_1$, $W_{22}$ becomes
\begin{equation}
\label{w22}
W_{22}(t)=\frac{1}{n}\int \frac{d^3p_1}{(2\pi)^3}\frac{p_{1z}^2}{E_1^2}\,
C_{22}\,,
\end{equation}
where $C_{22}$ does not contribute to the second integral in
Eq. (\ref{dotaniso}) due to particle number conservation in elastic
collisions. The same holds for the sum of $C_{23}$ and $C_{32}$ in chemical
equilibrium. Inserting the explicit expression of the collision term
\begin{eqnarray}
\label{c22}
C_{22}&=&\frac{1}{2E_1}\int d\Gamma_2\frac{1}{2!}
\int d\Gamma^{'}_1 d\Gamma^{'}_2\, f^{'}_1 f^{'}_2 
|{\cal M}_{1^{'}2^{'}\to 12}|^2 (2\pi)^4 
\delta^{(4)}(p^{'}_1+p^{'}_2-p_1-p_2)\nonumber \\
&&-\frac{1}{2E_1}\int d\Gamma_2\, f_1 f_2 \frac{1}{2!}
\int d\Gamma^{'}_1 d\Gamma^{'}_2\, |{\cal M}_{12\to 1^{'}2^{'}}|^2
(2\pi)^4 \delta^{(4)}(p_1+p_2-p^{'}_1-p^{'}_2)
\end{eqnarray}
($d\Gamma_i=d^3p_i/(2\pi)^3 2E_i$ for short) into Eq. (\ref{w22}) gives
two terms, which indicate the ``gain'' and ``loss'' in momentum
isotropization. 

The loss term is
\begin{equation}
\frac{1}{n} \int d\Gamma_1 d\Gamma_2
f_1 f_2 \frac{p_{1z}^2}{E_1^2}\,2s\,\sigma_{22}
=n \left \langle v_{\rm rel} \frac{p_{1z}^2}{E_1^2} \, \sigma_{22} 
\right \rangle_2\,,
\end{equation}
where
\begin{equation}
\sigma_{22}:=\frac{1}{2s} \frac{1}{2!} \int d\Gamma^{'}_1 
d\Gamma^{'}_2 \, |{\cal M}_{12\to 1^{'}2^{'}}|^2 \, 
(2\pi)^4 \delta^{(4)}(p_1+p_2-p^{'}_1-p^{'}_2)
\label{cs22}
\end{equation}
is the total cross section, $s$ is the invariant mass of the colliding
system, $v_{\rm rel}=s/2E_1E_2$ is the relative velocity, and
$\langle \ \rangle_2$ symbolizes an ensemble average over incoming
particle pairs. In BAMPS 
$f(x,p)=\sum_i \delta^{(3)}[\vec x-\vec x_i(t)] \delta^{(3)}(\vec p-\vec p_i)$
and we evaluate the averages $\langle \ \rangle_2$ in local
cells by running over all particle pairs in the cells. Each cell has
a small volume to ensure local collisions and has a sufficient
number of (test) particles to achieve adequate statistics.

The $W_{22}$'s gain term is $n\langle v_{\rm rel}\tilde \sigma_{22}\rangle_2$,
where
\begin{equation}
\tilde \sigma_{22}:=\frac{1}{2s} \frac{1}{2!} \int d\Gamma^{'}_1
d\Gamma^{'}_2 \, \frac{p^{'2}_{1z}}{E^{'2}_1}\,
|{\cal M}_{12\to 1^{'}2^{'}}|^2 \, (2\pi)^4 
\delta^{(4)}(p_1+p_2-p^{'}_1-p^{'}_2)\,,
\label{gain22}
\end{equation}
which, like Eq. (\ref{cs22}), is an integral over all possible states of
outgoing particles. Equation (\ref{gain22}) was obtained by exchanging
the primed and unprimed variables in Eq. (\ref{c22}). Except for
$p^{'2}_{1z}/E^{'2}_1$ all variables and functions in Eq. (\ref{gain22})
are Lorentz invariant. Particularly we find
$d\Gamma^{'}_1=d\Gamma^{'*}_1=d^3p^{'*}_1/(2\pi)^3 2E^{'*}_1
=d\Omega^*dE^{'*}_1 E^{'*}_1/2(2\pi)^3$, where $p^{'*}_1$ is the four-momentum
of an outgoing particle in the center-of-mass frame manifested by $p_1$
and $p_2$ of the incoming particles, and $\Omega^*$ denotes the solid
angle relative to the collision axis in the center-of-mass frame.
Integrating over $d\Gamma^{'}_2=d\Gamma^{'*}_2$ using the
four-dimensional $\delta$ function gives
\begin{equation}
\tilde \sigma_{22}=\int d\Omega^* \frac{d\sigma_{22}}{d\Omega^*} 
\, \frac{p^{'2}_{1z}}{E^{'2}_1}\,,
\end{equation}
where $p^{'}_{1z}$ and $E^{'}_1$ are the Lorentz transformed quantities
from $p^{'*}_1$ and, thus, functions of $\Omega^*$, $s$, and $\vec \beta$.
The $\vec \beta=(\vec p_1+\vec p_2)/(E_1+E_2)$ denotes the relative 
velocity of the center-of-mass frame of colliding particles to the
laboratory frame where $Q$ is defined.

We finally obtain
\begin{equation}
\label{trr22}
R^{\rm tr}_{22}= \frac{W_{22}}{Q_{\rm eq}-Q(t)}=\frac{1}{Q_{\rm eq}-Q(t)}\,
\left ( n \left \langle v_{\rm rel}\int d\Omega^* 
\frac{d\sigma_{22}}{d\Omega^*} 
\, \frac{p^{'2}_{1z}}{E^{'2}_1} \right \rangle_2
-n \left \langle v_{\rm rel} \frac{p_{1z}^2}{E_1^2} \, \sigma_{22}
\right \rangle_2 \right )\,,
\end{equation}
where the momentum isotropization gain and loss terms are clearly seen.
The relationship to the collision-angle distribution
is implicitly contained in $R^{\rm tr}_{22}$. When the collision rate is
defined as
\begin{equation}
\label{r22}
R_{22}=n\langle v_{\rm rel}\sigma_{22}\rangle_2\,,
\end{equation}
then we call $R^{\rm tr}_{22}$ the {\em transport collision rate} of elastic
scatterings.

The transport collision rate $R^{\rm tr}_{22}$ in Eq. (\ref{trr22}),
in general,
differs from $n\langle v_{\rm rel} \sigma^{\rm tr}_{22}\rangle_2$, where
$\sigma^{\rm tr}_{22}$ is defined in Eq. (\ref{tcs1}) or (\ref{tcs2}). They
match only if the laboratory frame is identical to the center-of-mass frame
of colliding particles. To demonstrate this we consider the special case
in which half of the particles move along the positive $z$ axis and the other
half of the particles move along the negative $z$ axis and all the particles
have the same energy $E$ such that
\begin{equation}
\label{special}
f(x,p) \propto \delta(p_x)\delta(p_y)\delta(p_z-E)+
\delta(p_x)\delta(p_y)\delta(p_z+E)\,.
\end{equation}
In this case the laboratory frame is the same as the center-of-mass frame
for every colliding pair. Thus $p^{'2}_{1z}/E^{'2}_1=\cos^2\theta^*$ and
$p^{2}_{1z}/E^{2}_1=1$. We then have
\begin{equation}
\label{trrate-cs22}
R^{\rm tr}_{22} =\frac{3}{2}\, n\langle v_{\rm rel} 
\sigma^{\rm tr}_{22}\rangle_2
\end{equation}
where $\sigma^{\rm tr}_{22}$ is given in Eq. (\ref{tcs1}). It is easy to verify
that Eq. (\ref{trrate-cs22}) does not depend on the direction of the initial
momentum. The only necessary conditions are that all the
particles move along the same (regardless of $\pm$ sign) direction and have
the same energy. Also, if $Q=\langle |p_z|/E\rangle$, $R^{\rm tr}_{22}$ will
be changed to
\begin{equation}
\label{trrate-cs22a}
R^{\rm tr}_{22} =2n\langle v_{\rm rel} \sigma^{\rm tr}_{22}\rangle_2\,,
\end{equation}
where $\sigma^{\rm tr}_{22}$ is given in Eq. (\ref{tcs2}).

The reason $R^{\rm tr}_{22}$ is called as the transport collision rate now
becomes obvious because Eq. (\ref{trr22}) is the generalization of the
simplified formula $n\sigma^{\rm tr}$, which is referred to in the literature
as the {\em transport collision rate} \cite{DG85,T94}.

To understand the physical meaning of the transport collision rate
it is reasonable to interpret $R^{\rm tr}_{22}$ as the rate per particle
at which particles experience elastic collisions to become isotropically
distributed in momentum space, because $R^{\rm tr}_{22}$ contributes to
momentum isotropization according to Eq. (\ref{separate}). For
ultrarelativistic particles the inverse of $R^{\rm tr}_{22}$ is the mean
path (or time) that particles should travel to become isotropic, and
$R_{22}/R^{\rm tr}_{22}$ is the average number of collisions, which each
particle needs to drive the particle system into isotropy in momentum space.

To confirm this interpretation we calculate $R^{\rm tr}_{22}$ assuming
that the collision angle is isotropically distributed. We then obtain
$R^{\rm tr}_{22}=R_{22}$ via Eq. (\ref{trrate-cs22}) or (\ref{trrate-cs22a})
for the special case (\ref{special}). This indicates that each particle
needs only one collision to drive the particle system into isotropy
in momentum space if the distribution of the collision angle is isotropic.
A more general case occurs during equilibration. The energy spectrum of
particles tends to be a Boltzmann distribution. Rarely found high-energy
particles need on the average more than one collision to become isotropic,
even if the distribution of the collision angle is isotropic. The reason
is that a particle with high energy always collides with
low-energy particles. The relative velocity of the center-of-mass frame to
the laboratory frame is large and, thus, the Lorentz boost has a strong
effect. In the laboratory frame deflection in the momentum of high-energy
particles is narrower in the forward direction. However, low-energy
particles move perpendicularly to their initial direction and, thus,
their momentum deflection is large. The averaged effect of the Lorentz
boost on the momentum isotropization is, however, nontrivial and
must be calculated numerically.

The above hinges on the assumption that the system is static. Expanding
systems are more complicated because
particles flow. Collisions not only deflect the particle momenta
but also force particles to flow. When we include the flow, which is
the particle drift contribution to momentum isotropization [see
Eq. (\ref{separate})], the momentum degradation of flowing particles
toward isotropy is slower ($\sim \theta_{\rm rel}$) than the inverse of
the transport collision rate, because the transport rate of particle drift
is negative in an expanding system.

\subsection{$R^{\rm tr}_{23}$ and $R^{\rm tr}_{32}$}
Compared with $W_{22}$ in Eq. (\ref{w22}), $W_{23}$ has an additional term
due to particle production
\begin{equation}
\label{w23}
W_{23}(t)=\frac{1}{n}\int \frac{d^3p_1}{(2\pi)^3}
\frac{p_{1z}^2}{E_1^2}\,C_{23}
-Q(t)\frac{1}{n}\int \frac{d^3p_1}{(2\pi)^3}\, C_{23}\,.
\end{equation}
Inserting the explicit formula
\begin{eqnarray}
C_{23}&&= \frac{1}{2E_1} \frac{1}{2!} \int d\Gamma_2 d\Gamma_3 
\frac{1}{2!} \int d\Gamma^{'}_1 d\Gamma^{'}_2 \,f^{'}_1 f^{'}_2
\, |{\cal M}_{1^{'}2^{'}\to 123}|^2 (2\pi)^4 
\delta^{(4)}(p^{'}_1+p^{'}_2-p_1-p_2-p_3)\nonumber \\
&&- \frac{1}{2E_1} \int d\Gamma_2 f_1 f_2 \frac{1}{3!}
\int d\Gamma^{'}_1 d\Gamma^{'}_2 d\Gamma^{'}_3 
|{\cal M}_{12\to 1^{'}2^{'}3^{'}}|^2 \, (2\pi)^4 
\delta^{(4)}(p_1+p_2-p^{'}_1-p^{'}_2-p^{'}_3)
\end{eqnarray}
into Eq. (\ref{w23}), we obtain
\begin{equation}
W_{23}(t)= \frac{3}{2} \, n \langle v_{\rm rel} \, \tilde \sigma_{23}\rangle_2
-n \left \langle v_{\rm rel} \frac{p^2_{1z}}{E^2_1}\, 
\sigma_{23}\right \rangle_2
-\frac{1}{2}\, Q(t)\, n \langle v_{\rm rel} \,\sigma_{23}\rangle_2 \,,
\label{w23_2}
\end{equation}
where
\begin{equation}
\tilde \sigma_{23}:=\frac{1}{2s} \frac{1}{3!} 
\int d\Gamma^{'}_1 d\Gamma^{'}_2 d\Gamma^{'}_3
\, \frac{p^{'2}_{1z}}{E^{'2}_1}\,
|{\cal M}_{12\to 1^{'}2^{'}3^{'}}|^2 \, (2\pi)^4
\delta^{(4)}(p_1+p_2-p^{'}_1-p^{'}_2-p^{'}_3)\,.
\label{gain23}
\end{equation}
The formula for $\sigma_{23}$ is just Eq. (\ref{gain23}), excluding
$p^{'2}_{1z}/E^{'2}_1$. The first two terms on the right-hand side of
Eq. (\ref{w23_2}), the sum of which is equal to the first term on the 
right-hand side of Eq. (\ref{w23}), have similar forms as those in 
Eq. (\ref{trr22}) [multiplying $Q_{\rm eq}-Q(t)$] for $W_{22}$.
The coefficients for the
momentum isotropization gain and loss terms, $3/2$ and $1$, indicate that
in a $2\to 3$ collision the ratio of the gained to the lost particle number
is $3/2$. The last term in Eq. (\ref{w23_2}) stems from pure particle
production. The coefficient for this term, $1/2$, comes from the sum of
the gain and loss terms in the particle production process. For a general
$M\to N$ collision the coefficients will be $N/M$, $1$, and $(N-M)/M$,
respectively. Assuming that
\begin{equation}
\label{decom1}
\left \langle v_{\rm rel} \frac{p^2_{1z}}{E^2_1}\, \sigma_{23}\right\rangle_2\,
\approx \,\left \langle \frac{p^2_{1z}}{E^2_1}\right\rangle\,
\langle v_{\rm rel}\, \sigma_{23}\rangle_2\\
= Q(t)\,\langle v_{\rm rel}\, \sigma_{23}\rangle_2 
\end{equation}
and then comparing $W_{23}$ in Eq. (\ref{w23_2}) to $W_{22}$ in 
Eq. (\ref{trr22}) [multiplying $Q_{\rm eq}-Q(t)$] we realize that
a $gg\to ggg$ collision is a factor of $3/2$ more efficient for momentum
isotropization than a $gg\to gg$ collision, when $\sigma_{22}=\sigma_{23}$
and $\tilde \sigma_{22}=\tilde \sigma_{23}$.
The physical reason is obvious: a $2\to 3$ collision brings one
more particle toward isotropy than a $2\to 2$ collision.

For the special distribution function (\ref{special}) we
find a relation between the transport collision rate and the transport
cross section (\ref{tcs1})
\begin{equation}
\label{trrate-cs23}
R^{\rm tr}_{23}=\frac{3}{2}\,\frac{3}{2}\, n\langle v_{\rm rel} 
\sigma^{\rm tr}_{23}\rangle_2\,.
\end{equation}
For scattering processes with isotropically distributed collision angles
one obtains $R^{\rm tr}_{23}=\frac{3}{2}\,R_{23}$ where
\begin{equation}
\label{r23}
R_{23}=n\langle v_{\rm rel} \sigma_{23}\rangle_2
\end{equation}
denotes the collision rate for a gluon undergoing $gg\to ggg$ collisions.
Bremsstrahlung effectively shortens the {\em mean transport path}
of particles that are becoming isotropic in momentum space.
Generally, in a $2\to N$ process
\begin{equation}
\label{trrate-cs2N}
R^{\rm tr}_{2N}=\frac{N}{2}\,\frac{3}{2}\, n\langle v_{\rm rel} 
\sigma^{\rm tr}_{2N}\rangle_2
\end{equation}
and the larger the number $N$, the stronger the effect. 

The final expression for $W_{32}$ (intermediate steps are analogous
to those for $W_{23}$, and $C_{32}$ is found in Ref. \cite{XG05}) is given by
\begin{equation}
W_{32}(t)= \frac{1}{3} n^2 \left \langle \frac{\tilde I_{32}}{8E_1E_2E_3}
\right \rangle_3 -\frac{1}{2} n^2 \left \langle \frac{p^2_{1z}}{E^2_1}
\frac{I_{32}}{8E_1E_2E_3} \right \rangle_3 +\frac{1}{6}\, Q(t)\, n^2
\left \langle \frac{I_{32}}{8E_1E_2E_3} \right \rangle_3\,,
\label{w32}
\end{equation}
where
\begin{equation}
\label{gain32}
\tilde I_{32}:=\frac{1}{2!} \int d\Gamma^{'}_1 d\Gamma^{'}_2
\, \frac{p^{'2}_{1z}}{E^{'2}_1}\,|{\cal M}_{123\to 1^{'}2^{'}}|^2
\, (2\pi)^4 \delta^{(4)}(p_1+p_2+p_3-p^{'}_1-p^{'}_2)\,.
\end{equation}
$I_{32}$ is just Eq. (\ref{gain32}), excluding $p^{'2}_{1z}/E^{'2}_1$.
$\langle \ \rangle_3$ denotes an ensemble average over triplets of
incoming particles.

Comparing $W_{23}$ to $W_{32}$, we see that the sum of the last terms 
in Eqs. (\ref{w23_2}) and (\ref{w32}) originates from the second term in 
Eq. (\ref{dotaniso}) but substituting $C_{23}+C_{32}$ in for
$\partial f/\partial t$ and it should be zero at chemical equilibrium.
We obtain
\begin{equation}
\label{chemeq}
n\langle v_{\rm rel}\,\sigma_{23}\rangle_2=\frac{1}{3}n^2
\left \langle \frac{I_{32}}{8E_1E_2E_3} \right \rangle_3
\end{equation}
or, equivalently, $R_{23}=\frac{2}{3}R_{32}$, where
\begin{equation}
\label{r32}
R_{32}=\frac{1}{2}n^2 \left \langle \frac{I_{32}}{8E_1E_2E_3}\right\rangle_3\,.
\end{equation}
From Eq. (\ref{chemeq}) we derived the collision rate of a gluon
experiencing $ggg\to gg$ collisions. Assuming that
\begin{equation}
\label{decom2}
\left \langle \frac{p^2_{1z}}{E^2_1}\frac{I_{32}}{8E_1E_2E_3}\right \rangle_3
\approx Q(t)\,\left \langle \frac{I_{32}}{8E_1E_2E_3}\right \rangle_3
\end{equation}
we finally have
\begin{eqnarray*}
W_{23}(t)&\approx& \frac{3}{2} ( n \langle v_{\rm rel} \, \tilde \sigma_{23}
\rangle_2 -Q(t)\, n \langle v_{\rm rel} \,\sigma_{23}\rangle_2 ) \\
W_{32}(t)&\approx& \frac{1}{3} n^2 \left \langle 
\frac{\tilde I_{32}}{8E_1E_2E_3} \right \rangle_3 -Q(t)\frac{1}{3} n^2
\left \langle \frac{I_{32}}{8E_1E_2E_3} \right\rangle_3\,.
\end{eqnarray*}
The expansion together with Eq. (\ref{chemeq}) leads to 
$W_{23}\approx \frac{3}{2} W_{32}$ and
$R^{\rm tr}_{23}\approx \frac{3}{2} R^{\rm tr}_{32}$ for chemical equilibrium.
Thus, a $2\to 3$ process should contribute more to kinetic equilibration
than a $3\to 2$ process because it brings one more particle toward
isotropy. If the system is out of chemical equilibrium, one expects
$R^{\rm tr}_{23}\approx \frac{3}{2} \frac{1}{\lambda_g} R^{\rm tr}_{32}$,
where the gluon fugacity $\lambda_g=1$ at chemical equilibrium.
In an undersaturated system ($\lambda_g < 1$), for instance, particle
production dominates and, therefore, $R^{\rm tr}_{23}$ is much larger
than $R^{\rm tr}_{32}$.

For the special case (\ref{special}) there is a direct relation
between the transport collision rate and transport cross section
[see Eqs. (\ref{trrate-cs22}) and (\ref{trrate-cs23})]. The same should be
self-evident for $R^{\rm tr}_{32}$ when detailed balance is considered
\begin{equation}
\label{trrate-cs32}
R^{\rm tr}_{32}\approx \frac{2}{3}\,\lambda_g\,R^{\rm tr}_{23}
=\frac{3}{2}\,\lambda_g\,n\langle v_{\rm rel} \sigma^{\rm tr}_{23}\rangle_2\,.
\end{equation}
If the distribution of the collision angle is isotropic,
$R^{\rm tr}_{32}\approx \lambda_g\,R_{23}= \frac{2}{3}\,R_{32}$, where
$\lambda_g=2R_{32}/3R_{23}$ is used. For a $N\to 2$ collision ($N > 2$)
\begin{equation}
\label{trrate-csN2}
R^{\rm tr}_{N2}\approx \frac{2}{N}\,\lambda_g\,R^{\rm tr}_{2N}
=\frac{3}{2}\,\lambda_g\,n\langle v_{\rm rel} \sigma^{\rm tr}_{2N}\rangle_2\,,
\end{equation}
which is not proportional to $N$ in contrast to $R^{\rm tr}_{2N}$ in
Eq. (\ref{trrate-cs2N}). For large $N$ a $2\to N$ process kinetically
equilibrates significantly more efficiently.

We summarize the main findings derived in this section:
\begin{itemize}
\item[(i)]
In Eq. (\ref{separate}) we showed that the relaxation rate of momentum
isotropization is a sum of the {\em transport rate} of particle drift and 
the {\em transport collision rates} of the various scattering processes.
\item[(ii)]
The transport rate of particle drift is negative for an expanding medium,
which means that the particle drift counteracts the momentum isotropization.
\item[(iii)]
The transport collision rates of the various interactions found in
Eqs. (\ref{trr22}), (\ref{w23_2}), and (\ref{w32}) [over $Q_{\rm eq}-Q(t)$]
have indirect but correctly implemented relationships with the
collision-angle distributions.
\item[(iv)]
$2\to N$ $(N>2)$ processes isotropize the momentum more efficiently than
elastic collisions or annihilation processes because the production
process brings more than two particles toward isotropy in momentum space.
\item[(v)]
The relations between the transport collision rates and the transport
cross sections for the special case in (\ref{special})
\begin{equation}
\label{sum1}
R^{\rm tr}_{22}=\frac{3}{2}\,n\langle v_{\rm rel} 
\sigma^{\rm tr}_{22}\rangle_2 \,,\,
R^{\rm tr}_{23}=\frac{3}{2}\,\frac{3}{2}\,
n\langle v_{\rm rel} \sigma^{\rm tr}_{23}\rangle_2\,,\,
R^{\rm tr}_{32} \approx \frac{3}{2}\,\lambda_g\,n\langle v_{\rm rel}
\sigma^{\rm tr}_{23}\rangle_2
\end{equation}
are found as long as $Q=\langle p^2_z/E^2\rangle$ and the transport cross
section is defined by Eq. (\ref{tcs1}). However, if
$Q=\langle |p_z|/E\rangle$ and the transport cross section is defined
by Eq. (\ref{tcs2}),
\begin{equation}
\label{sum1a}
R^{\rm tr}_{22}=2\,n\langle v_{\rm rel} \sigma^{\rm tr}_{22}\rangle_2 \,,\,
R^{\rm tr}_{23}=\frac{3}{2}\,2\,n\langle v_{\rm rel} 
\sigma^{\rm tr}_{23}\rangle_2\,,\,
R^{\rm tr}_{32} \approx 2\,\lambda_g\,n\langle v_{\rm rel}
\sigma^{\rm tr}_{23} \rangle_2\,.
\end{equation}
For the isotropic distribution of the collision angle we find
\begin{equation}
\label{sum2}
R^{\rm tr}_{22}=R_{22}\,,\qquad
R^{\rm tr}_{23}=\frac{3}{2}\,R_{23}\,,\qquad 
R^{\rm tr}_{32}\approx \frac{2}{3}\,R_{32}\,.
\end{equation}
\end{itemize}

\section{Results from the parton cascade calculations}
\label{sec:5}
In this section we present results on the gluon transport
rates in the central region of the expansion simulated by BAMPS.
We then compare the transport rates with
those obtained from the standard concept of the transport cross sections.
The potential dependence of the relaxation time on momentum is determined.

\subsection{Transport rate}
\begin{figure}[ht]
\centerline{\epsfysize=7cm \epsfbox{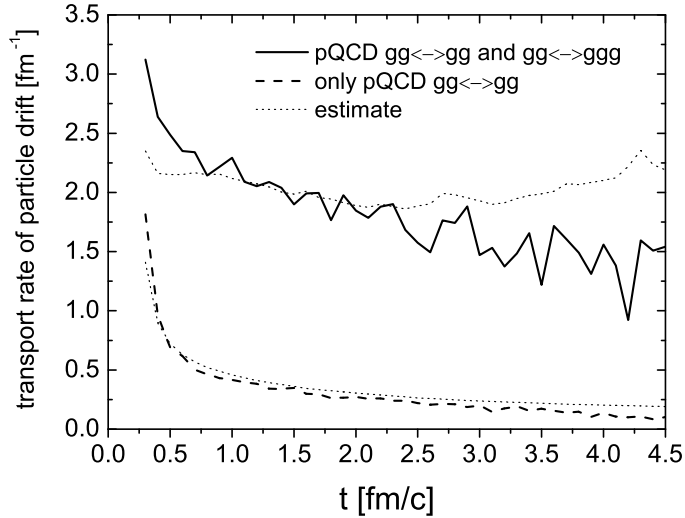}}
\caption{
Particle drift transport rate (multiplied by $-1$) from both elastic and
inelastic pQCD-scattering processes (solid curve), from elastic-only
pQCD-scattering processes (dashed curve), and from estimates in
Eq. (\ref{diff-2}) (dotted curves).
}
\label{tr-drift}
\end{figure}

In Fig. \ref{tr-drift} $-R^{\rm tr}_{\rm drift}$ is shown.
It cannot be computed by Eq. (\ref{diff}) because of strong numerical
uncertainties in calculating $\vec \nabla f$. Instead, it is obtained by
summing $\pm[Q(t)-p^2_z/E^2]$ over the particles, which come into (+ sign)
as well as leave (- sign) the central region within a time interval
of $0.1$ fm/c. This causes a large statistical fluctuation.
Comparing Fig. \ref{pze2} to Fig. \ref{tr-drift} we realize that the
larger the $Q=\langle p^2_z/E^2 \rangle$, the larger is the (negative) effect
of the particle drift on momentum isotropization. This confirms our
qualitative understanding outlined in Sec. \ref{sec:3}. The dotted
curves estimate the transport rate according to Eq. (\ref{diff-2}) assuming
a one-dimensional Bjorken boost-invariance expansion. $Q(t)$ and
$\langle p^4_z/E^4\rangle(t)$ come from the parton cascade.
At intermediate times our estimates nicely match the numerical results,
which indicates that the expansion follows a Bjorken expansion. Early
in the expansion the particle drift is stronger due to free streaming
caused by the initial conditions. Later on the expansion becomes
three-dimensional and particles begin to flow outward in the transverse
direction. The transverse drift of particles with large $p_T$ is then
similar to the longitudinal drift of particles with large $p_z\sim p_T$.
The net effect of the particle drift on momentum isotropization in
a three-dimensional expansion diminishes in comparison to a purely
longitudinal expansion, as demonstrated in Fig. \ref{tr-drift} with
the comparison of the numerical results to the estimations.

\begin{figure}[ht]
\centerline{\epsfysize=7cm \epsfbox{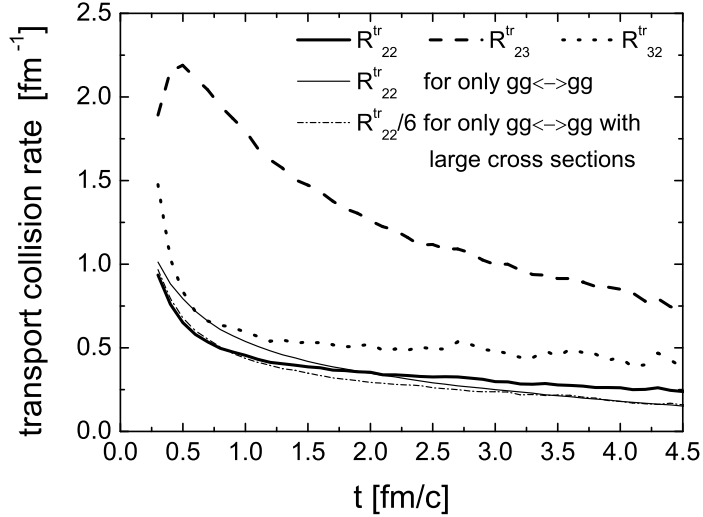}}
\caption{
Transport collision rates ($R^{\rm tr}_{22}$, $R^{\rm tr}_{23}$ and
$R^{\rm tr}_{32}$) from both elastic and inelastic pQCD-scattering
processes (thick solid, thick dashed, and thick dotted curve, respectively),
$R^{\rm tr}_{22}$ from elastic-only pQCD scatterings (thin solid curve)
and $R^{\rm tr}_{22}/6$ from elastic-only processes using artificially
large cross sections (thin dash-dotted curve).
}
\label{tr-coll}
\end{figure}

The numerical results for the transport collision rates are calculated
using the expressions (\ref{trr22}), (\ref{w23_2}), and (\ref{w32})
\{the last two are divided by $[Q_{\rm eq}-Q(t)]$\}, and are shown in
Fig. \ref{tr-coll}.
One realizes the dominance of the inelastic collisions in momentum
isotropization by computing the ratio 
$(R^{\rm tr}_{23}+R^{\rm tr}_{32})/R^{\rm tr}_{22}$, which is about $5$
throughout the entire evolution of the system. The ratio of 
$R^{\rm tr}_{23}$ to $R^{\rm tr}_{32}$ is always larger than $3/2$ but
nears $3/2$ late in the expansion. According to Eq. (\ref{trrate-cs32})
the system is undersaturated early on and eventually reaches chemical
equilibrium. When we compare $R^{\rm tr}_{22}$s obtained from the various
simulations, the difference is small, unlike for the cross sections shown in
Fig. \ref{cross-section}. The reason lies in the difference
in the evolution of the Debye sceening mass for the various simulations as
shown in Fig. \ref{md2}. A smaller Debye screening mass leads to a larger 
cross section but also a smaller collision angle. The former causes
more frequent collisions and, thus, speeds up equilibration, whereas
the latter causes inefficient momentum deflection and, thus, slows 
equilibration. Both contribute to the transport collision rate so that
it is not particularly sensitive to the Debye screening mass unlike
the total cross section. In Fig. \ref{cross-section} the total cross
sections for elastic collisions differ by a factor of $4-6$ between
elastic-only scatterings and those that include bremsstrahlung processes,
whereas the corresponding transport collision rates in Fig. \ref{tr-coll} 
are nearly identical.

The ratios of elastic+inelastic scatterings to elastic-only collisions
for the total transport collision rate and the transport rate of particle
drift are almost identical: the ratio increases from $4$ at $0.3$ fm/c 
to $9$ at $4.5$ fm/c. The inverse of the ratio of the momentum isotropization
time scales in the two simulations is also the same (see Fig. \ref{compare}).

Because the change in particle drift is a consequence of particle
collisions, one may expect that the momentum isotropization is
dependent only on the total transport collision rate. Gluon
kinetic equilibration would always look the same, if the total 
transport collision rate in every evolution was the same at every 
space-time point. The types of collision processes are not relevant,
although they are interesting in their own right. We have already shown
two examples of evolution of gluons in a central Au+Au collision at RHIC
energy. The total transport collision rate becomes on average a factor
of $6$ larger if pQCD bremsstrahlung processes are included. For another
evolution to have the same total transport collision rate as that
obtained when pQCD bremsstrahlung processes are included, elastic-only
scattering processes with larger cross sections, namely
$d\sigma_{22}/d\hat t=6\,d\sigma^{\rm pQCD}_{22}/d\hat t$, were used.
If the elastic pQCD cross sections obtained from the new simulation were
the same as those from elastic-only scatterings with pQCD cross sections
(see the dotted curve in Fig \ref{cross-section}), $6$ would be an 
appropriate prefactor. The dash-dotted curve in Fig. \ref{cross-section}
shows the elastic pQCD cross section calculated from the new simulation,
which is a factor of $2$ smaller than the dotted curve. Recalling that
the cross sections are dependent on the development of the Debye screening
mass and that the gluon evolution resembles free streaming for elastic-only
pQCD scatterings when the Debye screening mass decreases as $\sim t^{-1/2}$,
artificially large cross sections decrease the Debye screening mass
from $t^{-1/2}$ to $t^{-1/3}$ (see the dash-dotted curve in Fig. \ref{md2}),
which implies that the evolution of gluons for large cross sections
is a hydrodynamical expansion with a finite viscosity.

$R^{\rm tr}_{22}/6$ for large cross sections is depicted in Fig. \ref{tr-coll}
and is nearly the same as the transport collision rate for
standard pQCD cross sections, which proves that the transport collision
rate for elastic pQCD scatterings is not sensitive to the Debye screening
mass. Therefore, the total transport collision rates for elastic-only
collisions with large cross sections and for both elastic and inelastic
pQCD scatterings are nearly the same, which implies the same momentum
isotropization in both simulations. Comparing the time evolution of
the momentum isotropization (the solid versus the dash-dotted curve in
Fig. \ref{pze2}), we realize that momentum isotropization is indeed 
nearly the same. However, the total cross sections are very different
(see Fig. \ref{cross-section}). At $4.5$ fm/c, for instance, 
$\langle \sigma_{22}\rangle_2+\langle \sigma_{23}\rangle_2\approx 4$ mb
for elastic and inelastic scatterings, whereas
$\langle \sigma_{22}\rangle_2\approx 45$ mb for elastic-only collisions,
which is a factor of $12$ larger!

\begin{figure}[ht]
\centerline{\epsfysize=7cm \epsfbox{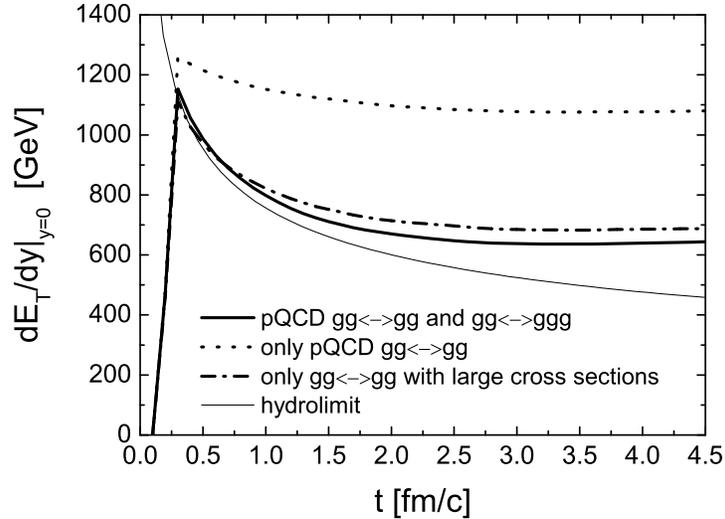}}
\caption{Time evolution of the transverse energy per unit momentum rapidity
at midrapidity for both elastic and inelastic pQCD scatterings (solid curve),
for elastic-only pQCD scatterings (dotted curve), for elastic-only
scatterings with large cross sections (dash-dotted curve), and the ideal
hydrolimit, $dE_T/dy|_{y=0}\sim t^{-1/3}$ (thin solid curve).
}
\label{et}
\end{figure}

Because kinetic equilibration and pressure buildup are related we also
expect that pressure buildup does not depend on the type of interactions.
Figure \ref{et} shows the time evolution of the transverse energy per
unit momentum rapidity at midrapidity.
The decrease in the transverse energy indicates that mechanical
work has been done by pressure gradients, which are built up during kinetic 
equilibration. From Fig. \ref{et} one realizes that the time
evolution of $dE_T/dy|_{y=0}$ obtained from elastic and inelastic scatterings
and from elastic-only scatterings with large cross sections are almost
identical. This indicates that the ongoing kinetic equilibration and the 
pressure gradients buildup are the same not only at the collision center
as already shown in Fig. \ref{pze2} but also at the central slice of the
expansion. There only the total transport collision rate matters, not
the detail of the interactions.

Whereas the decrease in $dE_T/dy|_{y=0}$ for elastic-only pQCD collisions
is very weak, which implies slow momentum isotropization, the decrease
in $dE_T/dy|_{y=0}$ in the other two is close to the ideal hydrodynamic
limit at least until $1.5$ fm/c. Later the expansion becomes
three-dimensional and gluons in the outer regions cease to interact when
the energy density decreases under the critical value of
$1 \,\mbox{GeV/fm}^3$. Therefore, the decrease in the transverse energy
slows so that the final value of $dE_T/dy|_{y=0}$ is about $650$ GeV,
which is comparable with RHIC data \cite{STAR04-2}.

Although the interaction details do not matter for kinetic equilibration
and pressure buildup, they do for chemical equilibration. Elastic
collisions conserve the absolute particle number and do not contribute
to chemical equilibration, whereas multiplication and annihilation processes
can drive systems toward chemical equilibrium. For the gluon evolution in
central Au+Au collisions the initial free streaming (or the off-shell
propagation) undersaturates the gluons (see Fig. \ref{fuga}). For
the pQCD bremsstrahlung processes chemical equilibrium is achieved by
producing gluons. This leads to a larger Debye screening mass than that
for elastic-only collisions (see Fig. \ref{md2}). Therefore, the elastic
pQCD cross section obtained from the simulation including the pQCD
bremsstrahlung processes is the smallest (see Fig. \ref{cross-section}).

\subsection{Mean free path, mean transport path, and relaxation time}
\begin{figure}[ht]
\centerline{\epsfysize=7cm \epsfbox{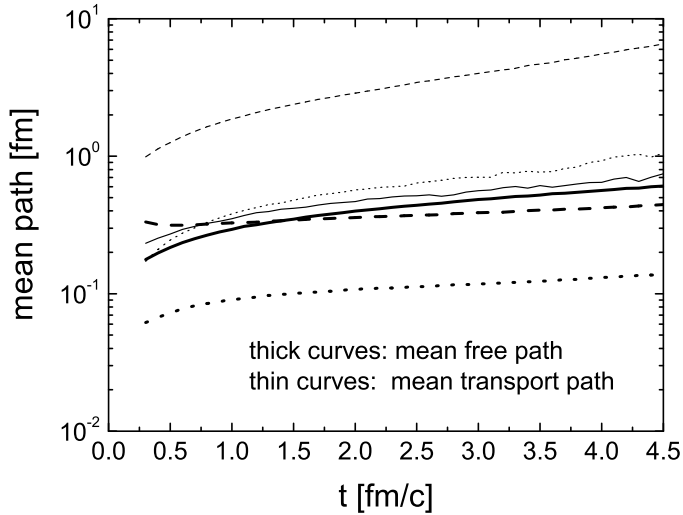}}
\caption{
Mean free path and mean transport path of gluons for
both elastic and inelastic pQCD scatterings (solid curves), for
elastic-only pQCD scatterings (dashed curves) and for elastic-only
collisions with artificially large cross sections (dotted curves).
}
\label{path}
\end{figure}

Figure \ref{path} shows the mean free path and the
{\em mean transport path} of gluons.
The {\em mean transport path} is defined as the inverse of
the total transport collision rate and it is the path needed for
gluons to reach isotropy in momentum space in a static medium.
From Fig. \ref{path} we see that the mean paths are all small early
on and increase throughout the course of the expansion when the system
becomes dilute. Comparing the mean free paths there is little
difference for the processes with and without pQCD bremsstrahlung;
however, the mean free path is much smaller when artificially large
cross sections are considered. We also see that the mean
transport path is larger than the mean free path for elastic-only
collisions, because elastic pQCD collisions have small-angle
scatterings and, therefore, do not isotropize the momentum efficiently.
However, when pQCD bremsstrahlung processes are included the mean
transport path and the mean free path are quite similar, so their
kinetic equilibration is efficient.

\begin{figure}[ht]
\centerline{\epsfysize=7cm \epsfbox{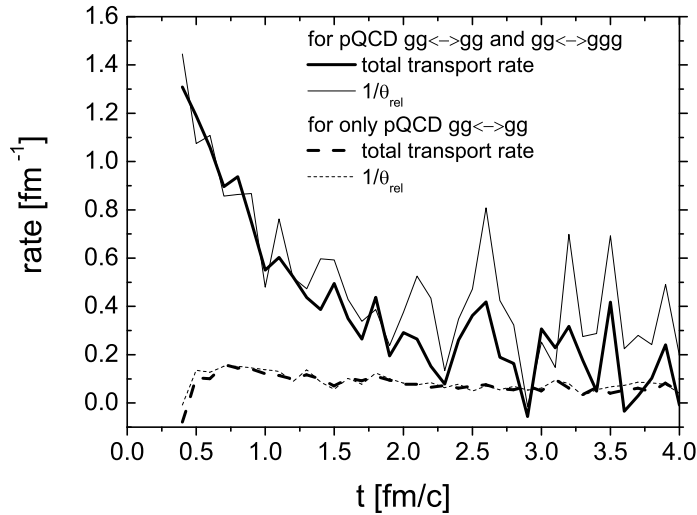}}
\caption{
Relaxation rate of momentum isotropization compared with
the total transport rate for both elastic and inelastic pQCD 
scatterings (thin and thick solid curves) and for elastic-only
pQCD scatterings (thin and thick dashed curves).
}
\label{compare}
\end{figure}

The relaxation rate of momentum isotropization $1/\theta_{\rm rel}(t)$
is calculated directly from Fig. \ref{pze2} using Eq. (\ref{relax2}) and
is shown in Fig. \ref{compare} in comparison with the total transport
rate $R^{\rm tr}_{\rm drift}+R^{\rm tr}_{22}+R^{\rm tr}_{23}+R^{\rm tr}_{32}$.
According to Eq. (\ref{separate}) they should be identical, which is
indeed seen within the numerical uncertainty. This indicates that
the transport rates were correctly extracted. For the first
$2$ fm/c of the gluon evolution for pQCD bremsstrahlung processes
the time scale of momentum isotropization is $1-2$ fm/c, which is 
about a factor of $5$ times larger than the mean free path
(see Fig. \ref{path}).

\subsection{Collision rate, transport collision rate,
and transport cross section}
Here we compare the collision rates, the transport collision rates and
the estimates using the transport cross sections with each other
concentrating on the results from the simulation with both elastic and
inelastic collisions.

\begin{figure}[ht]
\centerline{\epsfysize=7cm \epsfbox{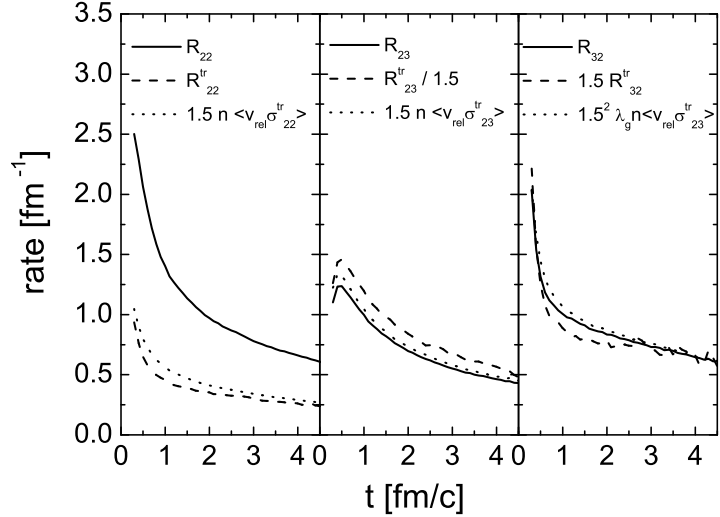}}
\caption{
Collision rate, transport collision rate, and
$n\langle v_{\rm rel}\,\sigma^{\rm tr}\rangle$ of the various collision
processes. The results are obtained from the simulation, including both
elastic and inelastic pQCD-scattering processes, and are depicted by
the solid, dashed, and dotted curves, respectively.
}
\label{rate-tr-tcs}
\end{figure}

Assuming Eq. (\ref{special}), the transport collision rates are directly
proportional to the transport cross sections [see Eqs. (\ref{sum1}),
(\ref{sum1a}), and (\ref{sum2})], which can be directly linked
to the collision angle distribution. To see how they differ from
the true transport collision rates we compare the transport
collision rates, $n\langle v_{\rm rel}\,\sigma^{\rm tr}\rangle$ and the
collision rates in Fig. \ref{rate-tr-tcs}.
Multiplication factors according to Eqs. (\ref{sum1}) and (\ref{sum2}) 
allow for more convenient comparisons. If the assumption (\ref{special})
is realistic, the curves according to the assumption and those for
the true transport collision rate are identical. If additionally the
collision-angle distribution is isotropic, all the curves in each case
lie on top of each other. The fugacity $\lambda_g$ in Eq. (\ref{sum1})
is calculated by $\lambda_g=n/n_{\rm eq}$, where
$n_{\rm eq}=16\,T^3/\pi^2$ is the gluon density at thermal equilibrium at
temperature $T\equiv\epsilon/3n$. The gluon density $n$ and energy
density $\epsilon$ are extracted from the parton cascade.

We first examine the rates for the elastic-scattering processes shown
on the left in Fig. \ref{rate-tr-tcs}, where there is only a small
difference between the true transport collision rate and the reduced
rate related to the transport cross section. The difference comes from
the Lorentz boost from the center-of-mass frame to the laboratory frame.
Furthemore, we see
that the transport collision rate is much smaller than the collision rate,
which is again due to the fact that the pQCD $gg\to gg$ scatterings are
small-angle scatterings and are not efficient for momentum isotropization. 

In the middle the rates for the $gg\to ggg$ bremsstrahlung processes are
shown where little difference is seen. The transport collision rate
divided by the kinematic factor $3/2$ is the largest rate, especially
over the collision rate. The kinematic factor for the assumption
(\ref{special}) is exactly $3/2$. However, in general it is only
approximately equal to $3/2$,
because the decomposition (\ref{decom1}) for $R^{\rm tr}_{23}$
[see $W^{\rm tr}_{23}$ in Eq. (\ref{w23_2})] is an approximation. The real
kinematic factor defined as $A$ may be larger than $3/2$, which would
lower $R^{\rm tr}_{23}/A$ below $R_{23}$. Even though
the difference between $\frac{2}{3} R^{\rm tr}_{23}$ and $R_{23}$ is
small, which indicates that the collision-angle distribution in
$gg\to ggg$ collisions is nearly isotropic. The same is also seen on
the right in Fig. \ref{rate-tr-tcs}, where the rates for $ggg\to gg$
are nearly identical.

Figure \ref{rate-tr-tcs} shows that the reduced transport
collision rates related to the transport cross sections do not
differ very much from the derived transport collision rates.
Generally, the Lorentz boost from the individual center-of-mass frame
to the laboratory frame does not lead to a big effect on momentum
isotropization. This is nontrivial. However, it
provides a basis to understand thermalization within multiparticle
reactions. Additionally, the transport collision rates derivation 
helps to obtain the {\em kinematic} factors in Eq. (\ref{sum1}), which
are essential in quantitative analyses but typically ignored in
the literature \cite{DG85,T94}.

Bremsstrahlung processes $gg\leftrightarrow ggg$ are suppressed by
the LPM effect, which occurs when a parton undergoes multiple scatters
with radiated gluons through a QCD medium (originally photons in the QED
medium). The interference of radiated gluons leads to suppression
of radiation of gluons with modes $(w,\vec k)$, where $w$ and $\vec k$
denote the gluon's energy and momentum. Heuristically, there
is no suppression for gluons with a {\it formation time} $\tau=w/k_T^2$
smaller than the mean free path. This is called the
Bethe-Heitler limit, where the gluon radiation induced at different
space-time points in the course of the propagation of a parton can
be considered as independent events. Events within the
Bethe-Heitler regime are included in BAMPS. Other gluon modes radiation
with coherent suppression completely drops out, which is the reason
for the $\Theta$ function in the matrix element in Eq. (\ref{m23}). 
Including these events speeds up thermalization; however, implementing
the coherent effect into a transport model where the Boltzmann equation
is solved remains a challenge.

Without implementing the LPM effect as a strict low momentum cutoff,
the matrix element for $gg\leftrightarrow ggg$ in Eq. (\ref{m23}) is
dominated by collinear bremsstrahlung, although it is suppressed
by the Debye screening mass as an infrared cutoff. Therefore, the larger
collision angle from $gg\leftrightarrow ggg$ processes in comparison
to elastic scatterings originates from the present implementation
of the LPM effect.

Because the angle of the radiated gluon relative to the
collision axis $\theta$ is related to the momentum rapidity $y$ by 
$\cos\theta=\tanh y$, which leads to  $\cosh y=1/\sin\theta$, the
effect of the $\Theta$ function in the matrix element (\ref{m23}),
$\Theta(k_{\perp}\Lambda_g-\cosh y)$, on the angular distribution of
the radiated gluon can be understood. For small transverse
momentum $k_{\perp}$, which the radiation favors, the rapidity $y$ is
small due to the $\Theta$ function if the mean free path
$\Lambda_g$ is small. This leads to large-angle radiation.
The larger $\Lambda_g$, the more small-angle bremsstrahlung
(with large $y$) occurs.

The $\Theta$ function results in a cutoff in the radiated gluon
phase space. The corresponding total cross section is found by
integrating the matrix element
\begin{equation}
\label{cs23}
\sigma_{gg\to ggg} \sim \int_0^{1/4} d\bar q_{\perp}^2
\int_{1/\bar \Lambda^2_g}^{1/4} d\bar k_{\perp}^2 \int_{-y_m}^{y_m} dy
\int_0^{\pi} d\phi \,
\frac{1}{({\bf \bar q}_{\perp}^2+{\bar m_D^2})^2}\,
 \frac{{\bf \bar q}_{\perp}^2}
{{\bf \bar k}_{\perp}^2 [({\bf \bar k}_{\perp}-{\bf \bar q}_{\perp})^2
+\bar m_D^2]}\, H(\bar q_{\perp}, \bar k_{\perp}, y, \phi)\,,
\end{equation}
where $\bar q_{\perp}^2=q_{\perp}^2/s$, $\bar k_{\perp}^2=k_{\perp}^2/s$,
$\bar \Lambda_g=\Lambda_g \sqrt{s}$, $\bar m_D^2=m_D^2/s$, $\phi$
is the angle between ${\bf k}_{\perp}$ and ${\bf q}_{\perp}$, and $H$
is a function of $\bar q_{\perp}$, $\bar k_{\perp}$, $y$, and
$\phi$. $H$ (found in Appendix D of Ref. \cite{XG05}) appears after
the integral over the radiating gluon momentum. 
In Eq. (\ref{cs23}) the lower cutoff for $k_{\perp}$ and the
upper (lower) cutoff $\pm y_m$ for $y$ can be seen, where $y_m$ is
the minimum among ${\rm arcosh}(\bar k_{\perp} \bar \Lambda_g)$ and 
${\rm arcosh}(1/2\bar k_{\perp})$ (see Appendix D of Ref. \cite{XG05}).
To obtain the radiated gluon angular distribution, which depends on
$\bar m_D^2$ and $\bar \Lambda_g$, one has to integrate over
$\bar k_{\perp}$, $\phi$ and $\bar q_{\perp}$ in Eq. (\ref{cs23}).
This is already done in Ref. \cite{XG05}.
The radiated gluon angular distribution and the
distributions of the other two gluons were depicted in Fig. 49
in Ref. \cite{XG05}, where $\bar m_D^2=0.05$ and $\bar \Lambda_g=4$.
The distributions are nearly isotropic. From the present BAMPS
calculation $\bar m_D^2 \approx 0.1$ and $\bar \Lambda_g \approx 3$
when pQCD bremsstrahlung is included. The value of $\bar m_D^2$ is almost
identical with the equilibrium value
$m_D^2/\langle s\rangle=4\alpha_s/3\pi=0.13$ for $\alpha_s=0.3$.
The smaller $\bar m_D^2$ found in Ref. \cite{XG05} is due to the slower
chemical equilibration, because the initial system
(using $p_0=2$ GeV) is more dilute than that used in this article
(using $p_0=1.4$ GeV). For larger $\bar m_D^2$ and smaller
$\bar \Lambda_g$ large-angle scatterings for $gg\to ggg$ are favored.

\begin{figure}[ht]
\centerline{\epsfysize=7cm \epsfbox{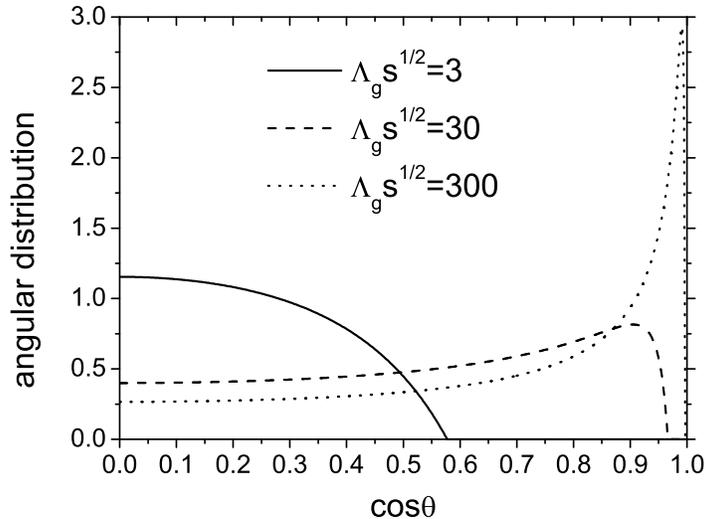}}
\caption{Radiated gluon angular distribution in the center-of-mass frame.
}
\label{trate-angle}
\end{figure}

Because the radiation is dominated by $1/{\bar k}_{\perp}^2$ we simplify
the matrix element (\ref{m23}) by eliminating the collinear term
$1/[({\bf \bar k}_{\perp}-{\bf \bar q}_{\perp})^2+\bar m_D^2]\sim 1/\bar m_D^2$
to see the effect of the LPM suppression (Bethe-Heitler regime)
on the radiated gluon angular distribution. Then the radiation can
be factorized
\begin{eqnarray}
\sigma_{gg\to ggg} \sim && 
\int_{1/\bar \Lambda_g}^{1/2} d\bar k_{\perp} \int_{-y_m}^{y_m} dy\,
\frac{1}{{\bar k}_{\perp}}=
\int_{-{\rm arcosh}\sqrt{\bar \Lambda_g/2}}^{{\rm arcosh}
\sqrt{\bar \Lambda_g/2}} dy 
\int_{\cosh y/\bar \Lambda_g}^{1/2\cosh y} d\bar k_{\perp} \,
\frac{1}{{\bar k}_{\perp}} \nonumber\\
&& =\int_{-{\rm arcosh}\sqrt{\bar \Lambda_g/2}}^{{\rm arcosh}
\sqrt{\bar \Lambda_g/2}} dy
\, \ln\frac{\bar \Lambda_g}{2\cosh^2 y}=
\int_{-\sqrt{1-2/\bar \Lambda_g}}^{\sqrt{1-2/\bar \Lambda_g}}\, du\,
\frac{\ln [ \bar \Lambda_g (1-u^2)/2]}{1-u^2}\,,
\end{eqnarray}
where $u=\cos\theta$. The integrand approximately represents the radiated
gluon angular distribution, which is bounded by 
$\pm\sqrt{1-2/\bar \Lambda_g}$. Figure \ref{trate-angle} shows the
distribution for various 
$\bar \Lambda_g=\Lambda_g \sqrt{s}\sim \Lambda_g T$.
The distributions are normalized and symmetric in $\cos\theta$.
The angular distribution is peaked in the forward direction only for
large $\bar \Lambda_g$ as in elastic scatterings. In BAMPS we find
$\bar \Lambda_g \approx 3$. Therefore, the radiated gluon angular
distribution according to Eq. (\ref{cs23}) is similar to 
$\Lambda_g \sqrt{s}=3$ in Fig. \ref{trate-angle}, which indicates that
large-angle radiation is favored.

\subsection{Dependence of the transport rate on 
the definition of $Q$}
We have already mentioned that the transport
rates depend on the definition of the degree of momentum isotropy $Q$.
In the previous subsections the numerical results for the transport rates
with $Q=\langle p_z^2/E^2\rangle$ were shown. But the dependence of
the numerical results of the transport rates on $Q$ (specially when
it is set to $Q=\langle |p_z|/E\rangle$) remains to be calculated.

Reasonable definitions of $Q$ must consider some kind of average of
the momentum spectra shown in Fig. \ref{distpxz}; thus, momentum 
isotropization time scales obtained from different prescriptions cannot
differ much from each other. Because the inverse of the total transport rate
is the momentum isotropization time scale, we do not expect any significant
dependence of the transport rate on the definition of $Q$.

Here we first compare the transport collision rates in the reduced
formulas (\ref{sum1}) for $Q=\langle p_z^2/E^2\rangle$ with those
in Eq. (\ref{sum1a}) for $Q=\langle |p_z|/E\rangle$, where transport
cross sections are defined by (\ref{tcs1}) and (\ref{tcs2}), respectively.
We have already shown that if the collsion angle is isotropically
distributed, the transport collision rates of a certain type of scattering
processes are the same, regardless of the definition of $Q$.
For small-angle scatterings one has $\sin^2\theta\approx \theta^2$ and
$1-\cos\theta \approx \theta^2/2$, and the transport collision rates for
$Q=\langle p_z^2/E^2\rangle$ are a factor of $1.5$ larger than those when
$Q=\langle |p_z|/E\rangle$. For large-angle scatterings
$\sin^2\theta \approx 1-\cos\theta$ and then the transport collision rates
for $Q=\langle p_z^2/E^2\rangle$ are in turn a factor of $3/4$ smaller
than those when $Q=\langle |p_z|/E\rangle$. The maximal relative difference
amounts to $50\%$. Because pQCD bremsstrahlung is the dominant
process in kinetic equilibration and the collision angle for that
process is roughly isotropic due to the LPM cutoff, the difference
in the transport collision rates due to different $Q$s should be minimal.

\begin{figure}[ht]
\centerline{\epsfysize=7cm \epsfbox{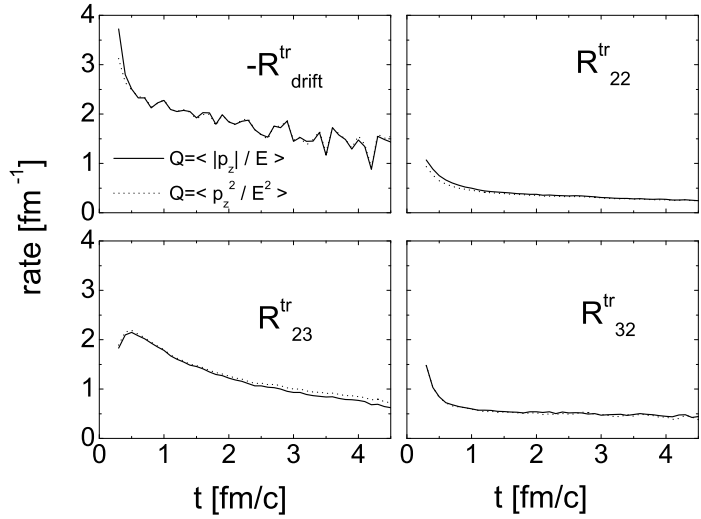}}
\caption{
Transport rates calculated with $Q=\langle |p_z|/E\rangle$ (solid curves)
and $Q=\langle p_z^2/E^2\rangle$ (dotted curves).
}
\label{trate-comp}
\end{figure}

Figure \ref{trate-comp} shows the numerical results for the transport
rates with the $Q$s defined above.
The differences are only small, which means that the transport rates
are not dependent on our choice in $Q$.

\subsection{Relaxation time $\tau_{\rm rel}$ in the relaxation time
approximation}
The collision term of the Boltzmann equation (\ref{boltzmann}) can be
written as
\begin{equation}
\label{rtapp}
C(x,p)=\frac{f_{\rm eq}(x,p)-f(x,p)}{\tau_{\rm rel}(x,p)}\,,
\end{equation}
which describes the relaxation of the particle-density function
$f(x,p)$ by using a space-time and momentum-dependent {\em relaxation time}
$\tau_{\rm rel}(x,p)$, where $\tau_{\rm rel}(x,p)$ is a functional of $f(x,p)$.
The underlying approximation in the so-called {\em relaxation time ansatz}
is the {\em ignorance} of the momenum dependence of the relaxation time,
i.e., $\tau_{\rm rel}(x,p)\approx \tau_{\rm rel}(x)$. Thus, $\tau_{\rm rel}(x)$
gives the time scale of the overall equilibration in absence of 
particle drift, which was used in \cite{baym,HK85,G91,HW96,W96,DG00,SS01}
to calculate the time scale of thermalization within various dynamical
scenarios of the expansion. It is crucial to see whether 
$\tau_{\rm rel}$ in the relaxation time approximation is equivalent to
the mean transport path, because the latter determines the momentum
isotropization time scale in a static system.

For kinetic equilibration we insert (\ref{rtapp}) into 
the time derivative of the momentum isotropization (\ref{dotaniso})
and obtain [by dividing ($Q_{\rm eq}-Q$)]
\begin{eqnarray}
R^{\rm tr}_{22}+R^{\rm tr}_{23}+R^{\rm tr}_{32}&=&\frac{1}{Q_{\rm eq}-Q}
\left ( \frac{1}{n}\int \frac{d^3p}{(2\pi)^3}\frac{p_z^2}{E^2} \,
\frac{f_{\rm eq}-f}{\tau_{\rm rel}}
-Q(t)\frac{1}{n}\int \frac{d^3p}{(2\pi)^3}
\frac{f_{\rm eq}-f}{\tau_{\rm rel}} \right ) \nonumber\\
&=&\frac{1}{Q_{\rm eq}-Q} \left ( 
\frac{n_{\rm eq}\,Q_{\rm eq}-n\,Q}{n\,\langle \tau_{\rm rel}\rangle_k}
-Q \frac{n_{\rm eq}-n}{n\,\langle \tau_{\rm rel}\rangle_c} \right )\,,
\label{tau0}
\end{eqnarray}
where $\langle \tau_{\rm rel}\rangle_k$ and $\langle \tau_{\rm rel}\rangle_c$
are defined as averaged quantities over the momentum, and the index $k$
denotes kinetic equilibration due to the convolution of angles ($p^2_z/E^2$)
in the first integration, whereas $c$ denotes chemical equilibration. 

$\langle \tau_{\rm rel}\rangle_c$ can also be calculated by integrating
the collision term of the Boltzmann equation over the momentum
\begin{equation}
\label{tauchem}
\int \frac{d^3p}{(2\pi)^3} \,(C_{22}+C_{23}+C_{32})
=\int \frac{d^3p}{(2\pi)^3} \,\frac{f_{\rm eq}-f}{\tau_{\rm rel}}
=\frac{n_{\rm eq}-n}{\langle \tau_{\rm rel}\rangle_c}\,,
\end{equation}
which is a simple ansatz for the relaxation time \cite{W96}.
The left-hand side of Eq. (\ref{tauchem}) is equal to
$n(R_{23}/2-R_{32}/3)$ if the explicit formulas of the collision
terms and the definition of the collision rates are applied.
We then obtain
\begin{equation}
\label{tau1}
\langle \tau_{\rm rel}\rangle_c=\frac{1/\lambda_g-1}{R_{23}/2-R_{32}/3}\,,
\end{equation}
where the gluon fugacity is $\lambda_g=n/n_{\rm eq}$.

Assuming that the relaxation time is independent of the momentum,
$\langle \tau_{\rm rel}\rangle_k$ and $\langle \tau_{\rm rel}\rangle_c$ 
become equal and one gets from Eq. (\ref{tau0})
\begin{equation}
\label{tau2}
\langle \tau_{\rm rel}\rangle_k=\frac{1}{\lambda_g}\,
\frac{1}{R^{\rm tr}_{22}+R^{\rm tr}_{23}+R^{\rm tr}_{32}}\,.
\end{equation}
In chemical equilibrium ($\lambda_g=1$) the relaxation time is 
equal to the inverse of the total transport collision rate or the
mean transport path. According to the relaxation time approximation
the right-hand sides of Eqs. (\ref{tau1}) and (\ref{tau2}) should be equal.
However, it is not clear.

Without assuming the relaxation time ansatz we can also calculate
the $\langle \tau_{\rm rel}\rangle_c$ and $\langle \tau_{\rm rel}\rangle_k$
using Eqs. (\ref{tau1}) and (\ref{tau0}), because all the collision rates
and the transport collision rates are known from numerical simulations.
If the two ``relaxation times'' differ much, one can conclude that
$\tau_{\rm rel}(x,p)$ is strongly momentum dependent and cannot serve
as a global quantity to determine the overall gluon thermalization 
time scale in ultrarelativistic heavy-ion collisions.

\begin{figure}[ht]
\centerline{\epsfysize=7cm \epsfbox{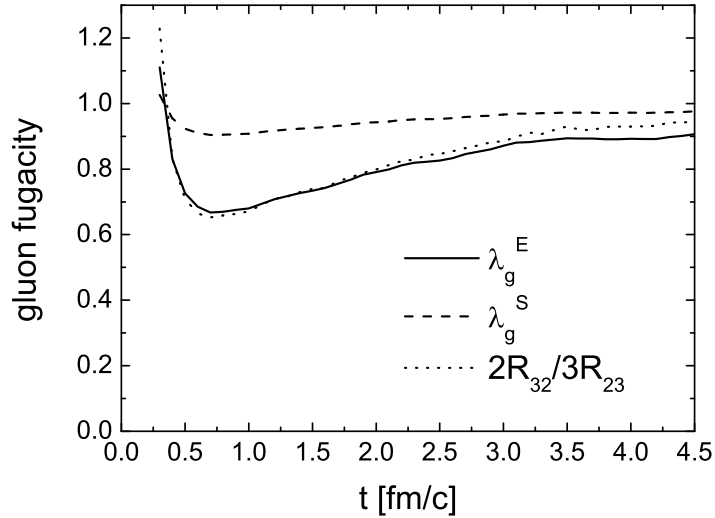}}
\caption{
Gluon fugacity. The solid, dashed, and dotted curve depict,
respectively, $\lambda^E_g$, $\lambda^S_g$, and $2R_{32}/3R_{23}$
obtained from the simulation, including elastic and inelastic
pQCD bremsstrahlung processes.
}
\label{fuga}
\end{figure}

Before we calculate $\langle \tau_{\rm rel}\rangle_c$
and $\langle \tau_{\rm rel}\rangle_k$, we need the equilibrium particle
density function $f_{\rm eq}(x,p)$. Because we neglect quantum effects
like gluon enhancement, $f_{\rm eq}(x,p)=\nu e^{-E/T}$ at the center of
the collision, where $\nu=16$ is the degeneracy of gluons. The temperature
$T$ can be found using $\epsilon_{\rm eq}=\epsilon$, which stems from 
energy conservation in sudden thermalization \cite{baym}.
However, the current particle density function could have
an exponential shape, $f=\lambda_g \, f_{\rm eq}$, if the kinetic
equilibration progressed quicker than chemical equilibration.
In this case one obtains $n=\lambda_g\,n_{\rm eq}$ as well as
$\epsilon=\lambda_g\,\epsilon_{\rm eq}$ and the temperature is then
$T=\epsilon_{\rm eq}/3n_{\rm eq}=\epsilon/3n$. This temperature is larger
(or smaller) than the previously defined temperature, if $\lambda_g$
is smaller (or larger) than $1$. The difference in these two local
temperatures leads to the difference in $\epsilon_{\rm eq}$, $n_{\rm eq}$,
and $\lambda_g$. Letting $S$ denote sudden thermalization and
$E$ thermalization that follows an exponential behavior, it is easy
to verify that
\begin{equation}
\label{fugacity}
\lambda^E_g=(\lambda^S_g)^4=\frac{27\pi^2}{16}\,\frac{n^4}{\epsilon^3}\,,
\end{equation}
where they differ by a power of $4$. The time evolution of the gluon
fugacities $\lambda^S_g$ and $\lambda^E_g$ obtained from BAMPS,
including pQCD bremsstrahlung processes is shown in Fig. \ref{fuga}.
Although the system of minijets is initially slightly
oversaturated, it becomes undersaturated due to a short period of
(quasi-)free streaming. The reason is obvious from Eq. (\ref{fugacity}),
when $n$ as well as $\epsilon$ decreases as $1/t$ in free streaming.
The decrease of $\lambda^E_g$ is roughly a factor of $4$ stronger than
that of $\lambda^S_g$. Whereas $\lambda^E_g$ increases and relaxes to $1$
later on, which indicates the ongoing chemical equilibration,
$\lambda^S_g \approx 1$ throughout the entire expansion, which implies
that the system is in chemical equilibrium. The difference between
$\lambda^S_g$ and $\lambda^E_g$ can be understood according to
Eq. (\ref{fugacity}), so
$(\lambda^E_g-1)\approx 4(\lambda^S_g-1)$ for $|\lambda^S_g-1|\ll 1$.

Physically, fugacity is a quantity that balances particle production
and annihilation. Therefore, the ratio of the annihilation rate $R_{32}$
to the production rate $R_{23}$ can serve as a quantitative measure of
fugacity, so, $2R_{32}/3R_{23}$ is shown in Fig. \ref{fuga}.
$2R_{32}/3R_{23}$ agrees well with $\lambda^E_g$, which implies that
$\lambda^E_g$ is an appropriate choice for the fugacity in this example.

\begin{figure}[ht]
\centerline{\epsfysize=7cm \epsfbox{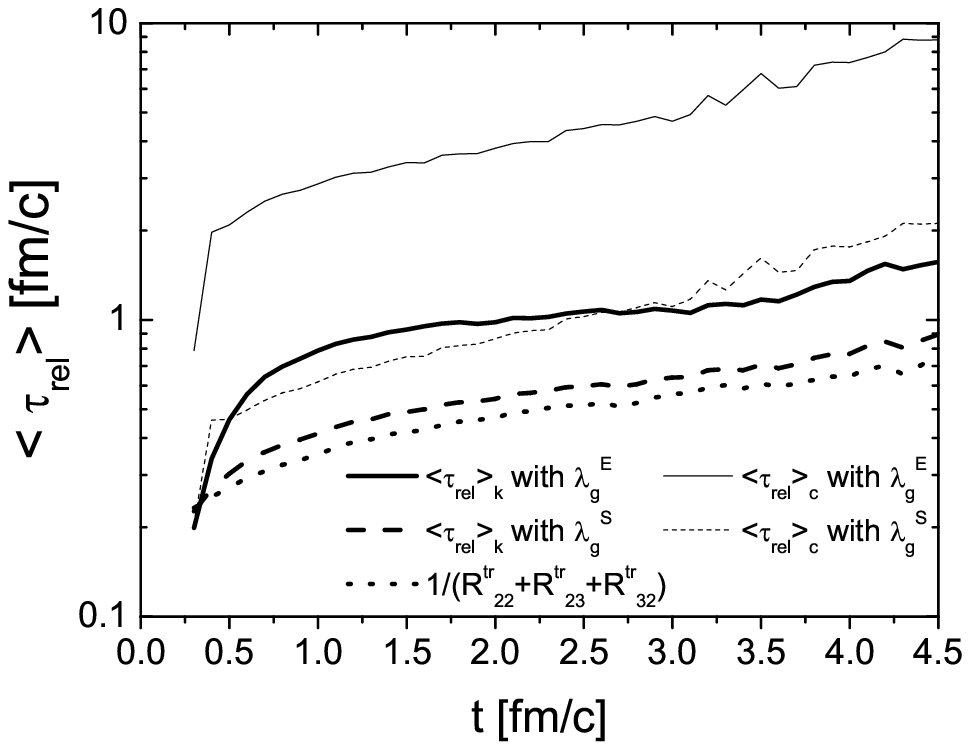}}
\caption{
Relaxation time. The thick (thin) solid curve depicts the momentum
averaged ``relaxation time'' $\langle \tau_{\rm rel}\rangle_k$ 
($\langle \tau_{\rm rel}\rangle_c$) using
the fugacity $\lambda^E_g$. The thick (thin) dashed curve depicts
$\langle \tau_{\rm rel}\rangle_k$ ($\langle \tau_{\rm rel}\rangle_c$) using
the fugacity
$\lambda^S_g$. The dotted curve shows again the mean transport path
(see Fig. \ref{path}). Results are obtained from the simulation,
including both elastic and inelastic pQCD-scattering processes.
}
\label{tau-rel}
\end{figure}

The exact momentum averaged ``relaxation times''
$\langle \tau_{\rm rel}\rangle_k$ and $\langle \tau_{\rm rel}\rangle_c$
according to Eqs. (\ref{tau0}) and (\ref{tau1}) are shown in 
Fig. \ref{tau-rel} by various gluon fugacities.
The mean transport path $1/(R^{\rm tr}_{22}+R^{\rm tr}_{23}+R^{\rm tr}_{32})$
is also depicted for comparison. Except for 
$\langle \tau_{\rm rel}\rangle_k$ with $\lambda^S_g$ all ``relaxation times''
are considerably larger than the mean transport path. When comparing
$\langle \tau_{\rm rel}\rangle_k$ to $\langle \tau_{\rm rel}\rangle_c$
with the same fugacities, one finds large differences, especially
for $\lambda^E_g$, where both ``relaxation times'' deviate by a factor
of $4$ to $6$. This implies that the relaxation time $\tau_{\rm rel}(x,p)$
in Eq. (\ref{rtapp}) indeed has a strong dependence on the momentum. 
Therefore, the applicability of the 
$\tau_{\rm rel}(x,p)\approx \tau_{\rm rel}(x)$ in studying gluon
thermalization in heavy-ion collisions is questionable. 

\section{Summary}
\label{summary}
Employing our recently developed parton cascade BAMPS and including 
inelastic pQCD bremsstrahlung processes we have introduced and
calculated the transport rate of gluon drift and the transport collision
rates of various scattering processes within relativistic kinetic theory.
We try to explain the observed fast equilibration of gluons within BAMPS
in theoretical terms.

We have shown that the derived transport rate of a certain process,
$R^{\rm tr}_{\rm drift}$, $R^{\rm tr}_{22}$, $R^{\rm tr}_{23}$,
or $R^{\rm tr}_{32}$, determines exactly the contribution of the process
to the defined momentum isotropization with $Q=\langle p^2_z/E^2 \rangle$ 
(or $Q=\langle |p_z|/E \rangle$). The total transport collision rate,
$R^{\rm tr}_{22}+R^{\rm tr}_{23}+R^{\rm tr}_{32}$, definitively describes
momentum isotropization, whereas the change of gluon drift is a consequence
of collision processes and $R^{\rm tr}_{\rm drift}$ is negative in
an expanding system. The inverse of the total transport rate,
$1/(R^{\rm tr}_{\rm drift}+R^{\rm tr}_{22}+R^{\rm tr}_{23}+R^{\rm tr}_{32})$,
gives the exact time scale of momentum isotropization $\theta_{\rm rel}$,
and is about $1$ fm/c from BAMPS for the gluon matter produced at RHIC.
It is also shown that the calculated transport rates are independent on
the definition of the degree of momentum isotropy $Q$.

The inclusion of quarks into BAMPS is straightforward, but it is not yet
completed. In the presence of quarks the Debye screening mass will be
slightly larger, which leads to a decrease in the cross sections for 
$gg\to gg$ and $gg\to ggg$ scatterings. This slightly slows thermalization.
However, further kinetic processes like $q+g\leftrightarrow q+g$,
$q+g\leftrightarrow q+g+g$ and $q+q\leftrightarrow q+q+g$ will speed up
gluon thermalization. In addition, the effect may be small, because the
initial quark density is tiny (30\%) in comparison to the gluon density.
Therefore, quark thermalization may be quite slow, but the study of
it is still in progress.

We also derived the reduced transport collision rates
related to the transport cross sections. They are only exact for
the special case when the center-of-mass frame of individual collisions
coincides with the laboratory frame where the medium is observed.
The deviations from the exact transport collision rates
stem from the effects of Lorentz boosts from the center-of-mass frame
to the laboratory frame. The numerical results show that the reduced rates
differ little from the exact ones for the evolution of gluons in 
relativistic heavy-ion collisions. Lorentz boosts
do not seem to have a major effect on momentum isotropization. Nevertheless,
the derivation of the transport collision rate helps to obtain the correct
kinematical factors in the reduced rates summarized in Eqs. (\ref{sum1})
and (\ref{sum1a}), which
have been typically ignored in the literature. For instance, our
analyses showed that a $2\to N$ $(N>2)$ production process is about
a factor of $(N-2)/2$ more efficient for momentum isotropization
than its back-reaction or an elastic-scattering process.

Using the numerical results of the transport collision rates for
the various scattering processes we have investigated the importance
of including the pQCD bremsstrahlung processes in thermalization.
The inclusion of the pQCD bremsstrahlung processes
and their back reactions, as implemented in BAMPS, increases the
efficiency for thermalization by a factor of $5$. Overall kinetic
equilibration and pressure buildup have a time scale of about
$1$ fm/c. The large efficiency stems
partly from the increase in the particle number for the final state 
of $gg\to ggg$ collisions but mainly from the almost isotropic
angular distribution in the bremsstrahlung process due to the effective
implementation of LPM suppression, which still needs to be further
developed.

Additionally, we have calculated the momentum averaged ``relaxation
times'' with various gluon fugacities and they differ significantly
from each other. This indicates a strong momentum dependence of the
gluon relaxation time $\tau_{\rm rel}$ in heavy-ion collisions. Thus,
using the standard relaxation time approximation of full kinetic
Boltzmann processes is questionable.

\acknowledgments
The authors thank B. M\"uller for the suggestion to analyze the thermalization
time scale within the kinetic parton cascade BAMPS. We are also
indebted to M. Gyulassy for enlightening discussions. The authors thank
J. Noronha-Hostler and B. Schenke for the careful reading of the manuscript.
C.G. thanks the Galileo Galilei Institute for Theoretical
Physics for its hospitality and the INFN for partial support
during the completion of this work.

\appendix

\section{Degree of momentum isotropy in the central region}
\label{app1}
The central region is described by a cylinder with
a radius of $r_b=1.5$ fm and a longitudinal extension of $2Z_b$.
The longitudinal boundary $Z_b=(\tanh \eta_b) \, t$ with $\eta_b=0.2$
increases linearly with time. Within the central region the degree of
momentum isotropy is defined by
\begin{eqnarray}
\label{iso_app}
Q(t):&=&\frac{1}{n}\int \frac{d^3p}{(2\pi)^3}
\frac{p_z^2}{E^2} \,\frac{1}{V} \int_0^{r_b}dr \, r \int_0^{2\pi} d\phi
\int_{-Z_b}^{Z_b} dz\, f(\vec x, t, p) \nonumber \\
&=&\frac{1}{n}\int \frac{d^3p}{(2\pi)^3}
\frac{p_z^2}{E^2} \,\frac{2}{V} \int_0^{r_b}dr \, r \int_0^{2\pi} d\phi
\int_0^{Z_b} dz\, f(\vec x, t, p)\,,
\end{eqnarray}
where 
\begin{equation}
\label{ndens_app}
n(t)=\int \frac{d^3p}{(2\pi)^3} \, \frac{2}{V} \int_0^{r_b}dr \, r 
\int_0^{2\pi} d\phi \int_0^{Z_b} dz\, f(\vec x, t, p)\,.
\end{equation}
$V=2\pi r_b^2 Z_b$ is the volume of the central region. The second
equation in Eq. (\ref{iso_app}) arises because of the symmetry of
$f(\vec x, t, p)$ under the $\vec x \to -\vec x$ exchange.  In the limit
$r_b\to 0$ and $\eta_b\to 0$ one has
\begin{equation}
\frac{1}{V}\int_0^{r_b} dr\, r \int_0^{2\pi} d\phi \int_{-Z_b}^{Z_b} dz\,
f(\vec x, t, p) \to f(\vec x, t, p) |_{\vec x=0}\,,
\end{equation}
which is the definition of the degree of the local momentum
isotropy in this limit [see Eq. (\ref{aniso}) in Sec. \ref{sec:4}].
The transport rates in this limit were already given in
Sec. \ref{sec:4}.

Taking the time derivative of $Q(t)$ yields
\begin{eqnarray}
\label{dotiso_app}
\dot Q(t)&=&\frac{1}{n}\int \frac{d^3p}{(2\pi)^3}\frac{p_z^2}{E^2} \,
\frac{2}{V} \int_0^{r_b} dr\, r \int_0^{2\pi} d\phi \left [
\int_0^{Z_b} dz\, \frac{\partial f}{\partial t}+\tanh \eta_b \,
f(\vec x_{\perp}, Z_b, t, p) \right ]\nonumber \\
&&-Q(t)\frac{1}{n}\int \frac{d^3p}{(2\pi)^3}
\frac{2}{V} \int_0^{r_b} dr\, r \int_0^{2\pi} d\phi \left [
\int_0^{Z_b} dz\, \frac{\partial f}{\partial t}
+\tanh \eta_b \,f(\vec x_{\perp}, Z_b, t, p) \right ]\,.
\end{eqnarray}
The second term in the brackets comes from the time derivative of
the boundary $Z_b$ and can be rewritten as
\begin{equation}
\tanh \eta_b \,f(\vec x_{\perp}, Z_b, t, p)=
\int_0^{Z_b} dz\, \frac{\tanh \eta_b}{Z_b} f(\vec x_{\perp}, Z_b, t, p)
=\int_0^{Z_b} dz\, \frac{1}{t} f(\vec x_{\perp}, Z_b, t, p)\,.
\end{equation}
The Taylor expansion of $f(\vec x_{\perp}, Z_b, t, p)$ at $\vec x$
to the first order yields
\begin{equation}
\label{taylor_app}
f(\vec x_{\perp}, Z_b, t, p) \simeq f(\vec x, t, p)
+\frac{\partial f(\vec x, t, p)}{\partial z} \, (Z_b-z)
\end{equation}
and we then obtain
\begin{eqnarray}
\label{dotiso1_app}
\dot Q(t)&\simeq&\frac{1}{n}\int \frac{d^3p}{(2\pi)^3}\frac{p_z^2}{E^2} \,
\frac{2}{V} \int_0^{r_b} dr\, r \int_0^{2\pi} d\phi 
\int_0^{Z_b} dz\, \left [ \frac{\partial f}{\partial t}
+\frac{(Z_b-z)}{t}\,\frac{\partial f}{\partial z}
\right ]\nonumber \\
&&-Q(t)\frac{1}{n}\int \frac{d^3p}{(2\pi)^3}
\frac{2}{V} \int_0^{r_b} dr\, r \int_0^{2\pi} d\phi
\int_0^{Z_b} dz\, \left [ \frac{\partial f}{\partial t}
+\frac{(Z_b-z)}{t}\,\frac{\partial f}{\partial z}
\right ]\,.
\end{eqnarray}
The $0$th-order contributions in Eq. (\ref{dotiso_app}) cancel due to
the definition of $Q(t)$. $(Z_b-z)/t$ expresses the relative velocity of
the boundary slice at $Z_b$ to the slice at $z$ where particles are
sitting. The second term in the brackets in Eq. (\ref{dotiso1_app}) appears
due to the increasing
longitudinal boundary of the central region, and it becomes smaller
when $Z_b\to 0$ (or $\eta_b\to 0$). According to the Boltzmann equation
(\ref{boltzmann}) the expression in the brackets can be written as
\begin{equation}
\frac{\partial f}{\partial t}+\frac{(Z_b-z)}{t}\,\frac{\partial f}{\partial z}
=-\frac{p_x}{E}\frac{\partial f}{\partial x}
-\frac{p_y}{E}\frac{\partial f}{\partial y}-\left (\frac{p_z}{E}
-\frac{Z_b-z}{t} \right ) \frac{\partial f}{\partial z}+C_{22}+C_{23}+C_{32}\,.
\end{equation}
The term $p_z/E-(Z_b-z)/t$ implies that only particles with longitudinal
velocity $p_z/E$ larger than the relative velocity of the boundary slice
$(Z_b-z)/t$ can drift out of the central region. This will be taken into
account when calculating the transport rate of particle drift within
the central region. The evaluations of the transport collision rates
$R^{\rm tr}_{22}$, $R^{\rm tr}_{23}$, and $R^{\rm tr}_{32}$ are more
straightforward. One only needs to replace the expressions derived in
the limit $r_b\to 0$ and $\eta_b\to 0$, which are already given in
Sec. \ref{sec:4}, by
\begin{equation}
R^{\rm tr}_i \to \frac{1}{V}\int_0^{r_b} dr\, r \int_0^{2\pi} d\phi
\int_{-Z_b}^{Z_b} dz\, R^{\rm tr}_i\,,
\end{equation}
where $i=22$, $23$, or $32$.


\end{document}